\documentclass[aps,floatfix,preprint,showpacs,preprintnumbers,nofootinbib,superscriptaddress,natbib]{revtex4}

\usepackage[titletoc]{appendix}
\usepackage{graphicx,float}
\usepackage{epsfig}		
\usepackage{dcolumn}
\usepackage{bm}
\usepackage{subfigure}

\usepackage[all]{xy}
\usepackage{amsmath,upgreek}
\usepackage{amssymb}

\usepackage{pdfpages}
\usepackage{color}
\usepackage{graphicx,epstopdf}
\usepackage[colorlinks=true,
 linkcolor=red,
 urlcolor=purple,
 citecolor=blue,hyperindex]{hyperref}

\def\0{\mbox{\tiny $0$}}
\def\1{\mbox{\tiny $1$}}
\def\2{\mbox{\tiny $2$}}
\def\3{\mbox{\tiny $3$}}
\def\4{\mbox{\tiny $4$}}
\def\5{\mbox{\tiny $5$}}
\def\6{\mbox{\tiny $6$}}
\def\7{\mbox{\tiny $7$}}
\def\8{\mbox{\tiny $8$}}
\def\9{\mbox{\tiny $9$}}

\def\f14{\mbox{\tiny $\frac{1}{4}$}}

\def\bb#1{\mbox{\footnotesize $(#1)$}}

\begin{document}
\title{Classical and statistical limits of the quantum singular oscillator}
\author{Caio Fernando e Silva}
\email{caiofernandosilva@df.ufscar.br}
\author{Alex E. Bernardini}
\email{alexeb@ufscar.br}
\affiliation{~Departamento de F\'{\i}sica, Universidade Federal de S\~ao Carlos, PO Box 676, 13565-905, S\~ao Carlos, SP, Brasil.}\date{\today}

\begin{abstract}
The classical boundaries of the quantum singular oscillator (SO) is addressed under Weyl-Wigner phase-space and Bohmian mechanics frameworks as to comparatively evaluate phase-space and configuration space quantum trajectories as well as to compute distorting quantum fluctuations.
For an engendered pure state \textit{quasi}-gaussian Wigner function that recovers the classical time evolution (at phase and configuration spaces), Bohmian trajectories are analytically obtained as to show how the SO energy and anharmonicity parameters drive the quantum regime through the so-called quantum force, which quantitatively distorts the recovered classical behavior. Extending the discussion of classical-quantum limits to a quantum statistical ensemble, the thermalized Wigner function and the corresponding Wigner currents are computed as to show how the temperature dependence affects the local quantum fluctuations. Considering that the level of quantum mixing is quantified by the quantum purity, the loss of information is quantified in terms of the temperature effects. Despite having contrasting phase-space flow profiles, two inequivalent quantum systems, namely the singular and the harmonic oscillators, besides reproducing stable classical limits, are shown to be statistically equivalent at thermal equilibrium, a fact that raises the SO non-linear system to a very particular category of quantum systems. 
\end{abstract}

\date{\today}
\maketitle

\section{Introduction}

Universal quantum descriptions which could circumstantially encompass larger sets of classical phenomena still undergo conceptual problems at the boundary of both classical and quantum domains \cite{Neumann,Zurek01}.
It includes, for instance, the measurement interpretation related to the wave-function collapse \cite{Zurek02,Bernardini13A}, the correspondence between uncertainty relations and observable quantities \cite{Catarina001,Stein,Bernardini13C,Bernardini13E}, the paradigmatic range of standard quantum mechanics \cite{Catarina,Bernardini13B,PhysicaA}, and even the system unitary evolution \cite{Eplus19,JCAP18} deeply related to the meaning of time \cite{Kiefer07,Bojowald,Ander12}, namely related to quantum cosmological scenarios \cite{Page83,Vilenkin94,Genovese,VilenBerto1,VilenBerto2}. 

Further steps toward an indefectible quantum mechanical framework are often supported by the classical mechanics, as a guidance for the subsidiary models. In this context, the fluid analogues of the information fluxes identified in the phase-space formulation of classical and quantum mechanics have provided a set of constraint equations which lead to novel quantifiers for quantumness (non-Liouvillian fluidity) \cite{EPL18,NovoPaper,NovoPaper02} sometimes given in terms of quantum decoherence, purity and entropy fluxes \cite{Steuernagel3,Donoso12,Liouvillian,NovoPaper,NovoPaper02}.

In particular, the Weyl-Wigner formalism \cite{Wigner,Moyal,Case} introduces a fluid analog of the phase-phase information flow, for which the Wigner function encodes all the information provided by the Schr\"odinger associated wave function or, more generally, the density matrix operator. 
As a {\em quasi}-probability distribution, the Wigner function can eventually exhibit some non-classical patterns since it can return negative values to the corresponding distribution function.
Pragmatically, the evolution of the Wigner function describes how the quantum phase-space ensembles evolve in time, and how quantum fluctuations quantitatively affect the Wigner flow \cite{Case}.
In the classical limit, the corresponding continuity equation is reduced to the Liouville equation for the classical probability distribution. The quantum fluctuations due to higher order derivatives of the quantum mechanical potential introduce distortions over a Liouvillian flow, and thus the non-Liouvillian property describes an unexplored set of quantum features, which can be associated to loss of information in the classical domain \cite{EPL18,JCAP18,NovoPaper}.
In the same context, an exhausting list of more enhanced frameworks, which includes either the Husimi $Q$ \cite{Husimi,Ballentine} or the Glauber-Sudarshan representation \cite{Glauber,Sudarshan,Carmichael,Callaway}, is mostly considered either to circumvent or to clear up the negative probability misunderstanding (cf. for instance the optical tomographic probability representation of quantum mechanics \cite{Amosov,Radon,Mancini}).
Otherwise, the fluid analogy and the analytical manipulability of the Weyl-Wigner formalism simplify its comparison with other quantum mechanical frameworks, as it can happen with respect to quantum back reaction formalism \cite{QBR} or even to the alternative fluid-like configuration space formulation provided by Bohmian quantum mechanics \cite{Bohm,Holland,Sanz,Durr}.

According to the Bohmian formulation \cite{Bohm,Holland}, quantum trajectories are obtained from the phase of the wave function in the polar form. Given that these trajectories are simply derived from the Schr\"{o}dinger associated continuity equation, they do not depend on any particular interpretation regarding their ontology. Instead, the ensemble trajectories provide an intuitive description of the micro-dynamics associated to the quantum system as it preserves the predictions from the quantum theory. Also, Bohmian trajectories exhibit constant probability patterns distinct from classical trajectories. Due to the so-called quantum force, which has nothing to do with the Newtonian one, classical-like motion depends on the shape of the probability distribution, and it is only recovered if the quantum force vanishes. 
Recently, the Bohmian mechanics has experienced a revival \cite{Santini,Ali,Neto} in the context of quantum cosmology theories, and the trajectories approach have been used to shed some light on the interpretation of their dynamical content, for which the classical solution breaks down.

Weyl-Wigner and Bohmian quantum mechanics are indeed complementary approaches: while one describes Liouville-like ensembles, the other one describes ensemble trajectories in the configuration space.
Due to its complementary aspects with respect to the Weyl-Wigner analysis, the complete Bohmian description of the singular oscillator (SO) quantum problem, which by the way supports several aspects of the Ho\v{r}ava-Lifshitz (HL) \cite{Hora09,Horava:2010zj,Blas:2009qj,Blas:2010hb} quantum cosmological models \cite{Mukohyama:2010xz,Zarro,Mukohyama:2009mz}, is evaluated along this paper. 

The classical SO corresponds to an anharmonic system driven by the harmonic oscillator potential modified by the addition of an inverse square contribution \cite{0001} which recovers the one-dimensional reduction of the three-dimensional radial equation for the Coulomb potential. It experiences an equivalent spectral decomposition to the one of the harmonic oscillator, as well as it can be analytically resolved by the Schr\"{o}dinger equation \cite{0003,0004,0005}.
Therefore, it works as an adaptable platform for investigating the limits and the interface between quantumness and non-linearities \cite{JCAP18, NovoPaper,NovoPaper02}.
Moreover, the SO exhibits thermal stable configurations which suggest it as an efficient frame for assessing the informational content of the quantum system, in particular, for discriminating the statistical properties of nonequivalent systems which, at thermal equilibrium, seems to match perfectly the same quantum information quantifiers related to quantum purity and to other related observables derived from partition functions.
More precisely, the thermodynamic descriptors associated to such a non-linear potential in the Wigner framework are closely related to the harmonic oscillator profile. Therefore, a more in-depth analysis into the phase-space versus the configuration profile of such anharmonic stable configurations obtained from the SO is evaluated.
In this context, it is expected that the complementary aspects devised by Wigner and Bohmian analysis can provide more enhanced views of this kind of non-linear dynamics.

Our manuscript is therefore organized as follows.
In Sec.~\ref{equasec2}, some grounds of Weyl-Wigner formalism are recovered in the context of re-introducing a simplifying dimensionless framework, through which one can report about the description of a {\em quasi}-gaussian pure state which reproduces the classical dynamics. In Sec.~\ref{equasec2B}, as to bring up the complementary information about the classical limits of this quantum system, Bohmian trajectories are analytically obtained and their quantum boundaries are discussed. In Sec.~\ref{equathermo}, the Wigner formalism is analytically extended from a pure state to a thermalized quantum statistical mixture. The phase-space quantum partition function and the purity quantifier are recovered from preliminary results \cite{Proceedings} as to drive the computation of the thermodynamic Wigner currents. In particular, the flow of information suggests how the classical regime is linked to the thermodynamic limit. Regarding the two different approaches, our conclusions are presented in Sec.~\ref{equaconclusions}.

\section{Singular oscillator {\em quasi}-gaussian pure state and the classical limit}\label{equasec2}

The quantum SO \cite{0003,0004,0005,NovoPaper}, which is a 1-dim harmonic oscillator modified by an inverse square contribution, can have its dynamics resumed by the driving potential written as
\begin{equation}
V(q) = \frac{m \omega^2}{2} q^2 + \frac{4\alpha^2 -1}{8m}\frac{\hbar ^2}{q^2} - \alpha \hbar \omega,
\end{equation}
where $m$ is the related mass and $\alpha$ is the quantifying anharmonic parameter.
The complete picture of the relevant physical aspects dictated by such a non-linear driving dynamics can be more conveniently obtained through a dimensionless description given in terms of a recasted energy scale $\hbar \omega$ \cite{NovoPaper}, for which the corresponding dimensionless Hamiltonian $\mathcal{H}(x,\,k) = (\hbar \omega)^{-1} H= k^2/2 + \mathcal{U}(x)$ is written in terms of dimensionless coordinate and momentum variables, $x = \left(m\,\omega\,\hbar^{-1}\right)^{1/2} q$ and $k = \left(m\,\omega\,\hbar\right)^{-1/2}p$, for a characteristic frequency $\omega$, such that the also recasted quantum mechanical potential is then written as $\mathcal{U}(x) = (\hbar \omega)^{-1} V\left(\left(m\,\omega\,\hbar^{-1}\right)^{-1/2}x\right)$ in order to return
\begin{equation}\label{equaqua14}
\mathcal{H}(k,\,x) = \frac{1}{2}\left\{k^2+ x^{2} + \frac{4 \alpha^2 -1}{4 x^{2}}-2\alpha \right\},
\end{equation}
for which, with the momentum operator identified by $k\equiv -i\,(d/dx)$, the time-independent Schr\"odinger equation reads
\begin{equation}\label{equaqua16}
\mathcal{H} \phi^{\alpha}_n(x) = \frac{1}{2}\left\{-\frac{d^{2}}{dx^{2}}+ x^{2} + \frac{4 \alpha^2 -1}{4 x^{2}}-2\alpha \right\} \phi^{\alpha}_n(x) = \varepsilon_n\,\phi^{\alpha}_n({x}),
\end{equation}
with energy eigenvalues $\varepsilon_n = 2n + 1$ (with $n$ integer) corresponding to the the energy spectrum $E_n = \hbar \omega (2n + 1)$, which also describes a harmonic oscillator of characteristic frequency given by $2 \omega$. The eigenfunctions are given by
\begin{equation}
\phi^{\alpha}_n(x) = 2^{{1}/{2}}\,\Theta(x) \, N_n^{(\alpha)}\, x^{\alpha + \frac{1}{2}}\,\exp(-x^2/2)\,L^{\alpha}_n(x^2),
\label{equaqua19}
\end{equation}
where $L^{\alpha}_n$ are the {\em associated Laguerre polynomials}, and $N_n^{(\alpha)}$ is the normalization constant given by
\begin{equation}
N_n^{(\alpha)} = \sqrt{\frac{n!}{\Gamma(n+\alpha+1)}}, 
\label{equaqua20}
\end{equation}
where $\Gamma(n) = (n-1)!$ is the {\em gamma function} and, finally, $\Theta(x)$ is the {\em heavyside function}, which constrains the solution to $0 < x < \infty$.
Through these preliminaries, an involving discussion of the semiclassical limit of the SO quantum system is given by the most general solution of the Schr\"{o}dinger equation, i.e. a time-dependent superposition, which can be written as
\begin{equation}
\mathcal{G}_\alpha (x, \tau) = \mathcal{N} \sum^\infty _ {n=0} c_n ^\alpha (\tau) \phi_n ^\alpha (x),
\end{equation}
for the time $\tau = 2\omega t$ also in a dimensionless form. A \textit{quasi}-gaussian wave packet is constructed by imposing $ c_n ^\alpha = \nu^n N_n ^{-1} (\alpha)\exp(i\,\tau/2)$ with $\nu = \exp(- \gamma + i \tau)$ and $\gamma > 0 $. One then has \[ \sum^\infty _ {n=0} \nu^n L^\alpha _n (x^2) = (1 - \nu)^{-\alpha - 1} \exp\left(\frac{x^2 \nu}{\nu - 1} \right), \]
and the quantum superposition, $\mathcal{G}_\alpha (x, \tau) $, is written as
\begin{equation}\label{equagaussiansuperposition}
\mathcal{G}_\alpha (x,\tau)= \mathcal{N} \Theta(x) x^{\alpha + \frac{1}{2}}(1-\nu)^{-(1+\alpha)}\exp\left[-\frac{1}{2}\left(\frac{1+\nu}{1-\nu}\right)x^2\right]\exp\left(\frac{i\, \tau}{2}\right),
\end{equation}
where 
\begin{equation}
\mathcal{N} = \left[\frac{(1-e^{-2\gamma})^{1+\alpha}}{2\Gamma(1+\alpha)}\right]^{1/2}.
\end{equation}
The probability density is then given by
\begin{equation}
\rvert \mathcal{G}_\alpha (x,\tau) \rvert^2 = 2\frac{u^{1+\alpha}}{\Gamma(1+\alpha)} \Theta(x) x^{1+2\alpha} \exp(- u x^2), \label{equadensidadeHL}
\end{equation}
where
\begin{equation}\label{equaeqaa}
u \equiv u (\tau) = \frac{\sinh(\gamma)}{\cosh(\gamma) - \cos(\tau + \varphi) }
\end{equation} 
and $\varphi $ is arbitrary. 

As to obtain a more enhanced interpretation of the dynamical aspects of the above SO quantum system, a description that can be extended to phase-space involves the computation of the so-called Wigner function associated to the above obtained {\em quasi}-gaussian distribution.
In this context, the quantum phase-space setup is generically achieved by the Weyl transform of a quantum operator $\hat{O}$ \cite{Case},
\begin{eqnarray}\label{equaweyl}
O^W (q,p) = &&2\int dw \exp(2 \,i \,p \, w /\hbar) \langle q + w\rvert \hat{O} \rvert q - w \rangle \nonumber \\
&& 2\int ds \exp(- 2 \,i \,q \, s/ \hbar) \langle p - s\rvert \hat{O} \rvert p + s \rangle,
\end{eqnarray}
which maps an arbitrary operator into a real function of the phase-space coordinates $q$ and $p$.
When the Weyl transform (\ref{equaweyl}) is applied onto the density operator in the configuration space, $\hat{\rho} = \rvert \psi \rangle \langle \psi \rvert$, for a generic quantum state $\psi$, the Wigner function \cite{Wigner}
has its inception identified by 
\begin{equation}
\mbox{Tr}[\hat{\rho}\hat{O}] = \langle O \rangle = \int dq \int dp \, W(q,p) \, O^W (q,p),
\end{equation}
so that
\begin{equation}
\mbox{Tr}[\hat{\rho}]= \int dq \int dp \, W(q,p) = 1,
\end{equation}
which guarantees the quantum unitarity properties.
The corresponding (dimensionless) Wigner function obtained from $\mathcal{G}_\alpha (x,\tau)$ is evaluated from Eq.~\eqref{equagaussiansuperposition} as to give
\begin{equation}
\mathcal{W}^{\alpha}(x, \, k;\,\tau) = \pi^{-1} \int^{+\infty}_{-\infty} \hspace{-.15cm}dy\,\exp{\left(2\, i \, k \,y\right)}\,\mathcal{G}_\alpha(x - y;\,\tau)\,\mathcal{G}_\alpha^{\ast}(x + y;\,\tau),\quad \mbox{with $y = \left(m\,\omega\,\hbar^{-1}\right)^{1/2} w$}
\end{equation}
which returns
\begin{eqnarray}
\label{equaeqn27}
\mathcal{W}^{\alpha}(x,\,k;\,\tau_{}) &=& \frac{2\, u^{1+\alpha}}{\pi\,\Gamma(1+\alpha)}\,\int_{-\infty}^{+\infty}dy\,\Theta(x+y)\Theta(x-y)\,(x^2-y^2)^{\frac{1}{2}+\alpha}\nonumber\\
&&\qquad\qquad\qquad \exp\left(-u(x^2+y^2)\right)\,\exp\left(2\,i\,y(k + v\, x)\right)\\
&=& \frac{2\,u^{1+\alpha}}{\pi\,\Gamma(1+\alpha)}\,x^{2(1+\alpha)}
\exp\left(-u x^2\right)\,\nonumber\\
&&\qquad \int_{-1}^{+1}ds\, (1-s^2)^{\frac{1}{2}+\alpha}\, \exp\left(- u \,x^2\,s^2\right)\,\exp\left(2\,i\,x\,s (k + v \, x)\right),\nonumber
\end{eqnarray}
for $y = x\,s$ and
\begin{equation}
v = - \frac{\sin(\tau)}{\cosh(\gamma) - \cos(\tau)},
\end{equation}
and which (as can be seen from Appendix I) obeys normalization and purity constraints (cf. Ref.~\cite{JCAP18}).

The above analytical structure has been investigated in the context of the HL cosmology, for which the Wigner currents, which depict the dynamical behavior of the Wigner function, have been exactly computed, as to include all phase-space features in the associated quantum description \cite{JCAP18}.
Moreover, it has been noticed that the correspondence between quantum and classical cosmological regimes could be identified and quantified as to have probabilistically dominant regions of the Wigner function following the classical trajectory of the HL universe in an exact (statistical) way \cite{JCAP18}.
In that case, the classical trajectory is mapped by the SO Hamiltonian, where its classical version 
is identified by $\mathcal{H} \rightarrow \mathcal{H}_\mathcal{C}$, with the energy parameter, $\mathcal{H}_\mathcal{C} = \varepsilon$, for which the Poisson brackets give
\begin{eqnarray}
\dot{k}_{_{\mathcal{C}}} &=& \frac{1}{2}\,\{k_{_{\mathcal{C}}},\,\mathcal{H}\}_{\mbox{\tiny PB}} = -\frac{1}{2} \frac{\partial \mathcal{H}}{\partial x_{_{\mathcal{C}}}} = -\frac{1}{2}\left(x_{_{\mathcal{C}}}- \frac{4 \alpha^2 -1}{4 x_{_{\mathcal{C}}}^{3}}\right),
\label{equaqua42A}
\\
 \dot{x}_{_{\mathcal{C}}} &=& \frac{1}{2}\,\{x_{_{\mathcal{C}}},\,\mathcal{H}\}_{\mbox{\tiny PB}} = +\frac{1}{2} \frac{\partial \mathcal{H}}{\partial k_{_{\mathcal{C}}}} = \frac{1}{2} k_{_{\mathcal{C}}},
\label{equaqua42}
\end{eqnarray}
where ``{\em dots}'' stand for $\tau$ derivatives, ``$\mathcal{C}$'' denotes the classical coordinates, and factors $1/2$ come from $\tau = 2 \omega t$. From Eqs.~(\ref{equaqua42A})-(\ref{equaqua42}), one has
\begin{eqnarray}
x_{_{\mathcal{C}}}\bb{\tau_{}} &=& \sqrt{\alpha +\varepsilon + \Delta \cos(\tau +\vartheta)}, \label{equaqua43}\\
k_{_{\mathcal{C}}}\bb{\tau_{}} &=& \frac{\Delta \sin(\tau +\vartheta)}{\sqrt{\alpha +\varepsilon + \Delta \cos(\tau +\vartheta)}},
\label{equaqua44}
\end{eqnarray}
constrained to $x_{_{\mathcal{C}}} > 0$ with $\Delta = \sqrt{\varepsilon^2 + 2\alpha\varepsilon +1/4}$ and $\vartheta$ arbitrary. When the above solutions are replaced into Eq.~(\ref{equaeqn27}) as to give $\mathcal{W}^{\alpha}(x_{_{\mathcal{C}}}\bb{\tau_{}},\,k_{_{\mathcal{C}}}\bb{\tau_{}};0)$, for
\begin{equation}\label{equagamma}
\gamma = \text{arccosh} \left( \frac{\alpha + \varepsilon}{ \Delta} \right),
\end{equation}
one has noticed that $\mathcal{W}^{\alpha}(x_{_{\mathcal{C}}}\bb{\tau_{}},\,k_{_{\mathcal{C}}}\bb{\tau_{}};0) = \mathcal{W}^{\alpha}(x,\,k;\,\tau_{})$, which is consistent with the above mentioned quantum to classical correspondence.
Moreover, the physical relevance of such a ({\em quasi})gaussian wave packet treatment is also usually associated to the construction of minimum uncertainty states. However, peaked wave packets spread out according to Schr\"{o}dinger and Wigner flow equations (cf. Appendix II), and a precise definition of the semiclassical limit is lost. As it shall be discussed in the following section, quantum particle Bohmian trajectories can add some insight into the system micro-dynamics, as to fill the blanks of the phase-space analysis.

\section{Classical limit revisited by the Bohmian approach}\label{equasec2B}
 
The {\em quasi}-gaussian superposition embedded into the Wigner phase-space formalism suggests an interesting path to investigate the deviation of quantum from classical trajectories. Alternatively, the Bohmian picture of quantum mechanics is achieved by supplementing the wave function with particle trajectories. Bohmian trajectories describe how a given point in the configuration space evolves according to a local velocity field, as opposed to a momentum distribution in the Wigner-Liouville approach. The single-valued Bohmian momentum is indeed more intuitive in the hydrodynamic picture of quantum mechanics \cite{Burghardt}, given that it traces back to the very foundation of the theory \cite{Madelung}. Once that one interprets the meaning of Bohmian trajectories, paths of constant probability are identified -- according to the quantum equilibrium hypothesis (see Sec. 2.5 of \cite{Allori}) -- so as to yield the usual statistical distribution of physical observables, which is fundamentally different from classical trajectories. 

For a generic wave function in the polar form, $\psi(q,t) = \rho^{1/2}\exp(i S /\hbar)$, where $\rho$ is the probability density and $S$ is the quantum phase, the one-dimensional Schr\"{o}dinger equation is read as \cite{Sanz}
\begin{equation}\label{equacontinuitybohm}
 \frac{\partial \rho}{\partial t} + \mbox{\boldmath$\mbox{\boldmath$\nabla$}$} \cdot \left( \rho \,\frac{\mbox{\boldmath$\mbox{\boldmath$\nabla$}$} S}{m} \right) = 0,
 \end{equation} 
\begin{equation}\label{equaQHJ}
\frac{\partial S}{\partial t} + \frac{(\mbox{\boldmath$\nabla$} S)^2}{2m} + V + Q = 0, \quad \text{for} \quad Q = - \frac{\hbar^2}{2m}\frac{\mbox{\boldmath$\nabla$}^2 \rho ^{1/2}}{\rho^{1/2}},
\end{equation}
where, from this point, $\mbox{\boldmath$\nabla$}$ shall be reduced to the one-dimensional notation $\mbox{\boldmath$\nabla$}\sim \partial_q$. Eq.~\eqref{equacontinuitybohm} is the continuity equation, and Eq.~\eqref{equaQHJ} is the quantum Hamilton-Jacobi equation, which introduces a force-like field from the novel potential term $Q$, as $F_q = - \partial_q Q$. Therefore, classical-like motion is expected for vanishing $\partial_q Q$ values. However, as already mentioned, the continuity equation constrains the local momentum to $p(q,t) = \partial_q S$, and thus the Newton's second law is not deducible from Eq.~\eqref{equaQHJ} \cite{Holland,Durr}. Nevertheless, one can still investigate single phase-space trajectories, for which the classical-quantum analogy becomes more intuitive. 

Moving to the dimensionless wave function in the polar form $\phi(x, \tau) = R^{1/2}\exp(i \mathcal{S})$ where $\hbar \mathcal{S}(x, \tau) = S(\,(m \omega \hbar^{-1})^{-1/2}x, t)$ and $(m \omega \hbar^{-1})^{1/2} R(x,\tau) =  \rho(\,(m \omega \hbar^{-1})^{-1/2}x, t)$, one can identify $\phi(x, \tau)$ with the stationary state from Eq.~\eqref{equaqua19}, $\phi^\alpha _n$. In this case, the equations for $R^\alpha _n (x,\tau)\equiv R$ and $\mathcal{S}^\alpha _n(x, \tau)\equiv \mathcal{S}$ are given by
\begin{equation}
 \frac{\partial R}{\partial \tau} + \partial_x \left( R \,\frac{\partial_x \mathcal{S}}{2} \right) = 0,
 \end{equation} 
\begin{equation}
2 \frac{\partial \mathcal{S}}{\partial \tau} + \frac{(\partial_x \mathcal{S})^2}{2} + \mathcal{U} + \mathcal{Q} = 0, \quad \text{for} \quad \mathcal{Q} = - \frac{1}{2}\frac{\partial_x^2 R^{1/2}}{R^{1/2}},
\end{equation}
where the extra factor $2$ sets the proper correspondence between the natural frequency of the system $\omega$ and the frequency described by the Hamiltonian, $\omega'= 2\omega$. Stationary states have position-independent $\mathcal{\mathcal{S}}$, and thus $k_n = \partial_x \mathcal{S} /2 $ vanishes -- that is, stationary wave functions provide stationary trajectories\footnote{Such a nonintuitive feature was extensively discussed in \cite{Bohm}.}. 
Turning back to the SO {\em quasi}-gaussian superposition from Eq.~\eqref{equagaussiansuperposition} cast in the form of $\mathcal{G}_\alpha = {R_\alpha} ^{1/2}\exp(i \mathcal{S_\alpha})$, after some simple algebraic manipulations\footnote{Notice that $\exp\left[-\frac{x^2}{2}\left(\frac{1+\nu}{1-\nu}\right)\right] = \exp\left[-\frac{x^2u}{2} + i \frac{x^2 v}{2}\right]$ and $\left(\frac{1}{1-\nu}\right)^{1 + \alpha} = \left(\frac{1+ u - iv}{2}\right)^{1 + \alpha}$, where the numerator inside the brackets can be written as $r \exp(i \theta)$ for $r = \sqrt{(1 +u)^2 + v^2} \quad \text{and} \quad \theta = \arctan\left(\frac{-v}{1 + u}\right).$} one finds
\begin{equation}\label{equaphase}
\mathcal{S_\alpha}(x,\tau) = \arctan \left(\frac{v}{1 + u} \right)(1 + \alpha) - \frac{v x^2}{2} - \frac{ \tau}{2},
\end{equation}
where again 
\begin{equation}
 u = \frac{\sinh(\gamma)}{\cosh(\gamma) - \cos(\tau)} \quad \text{and} \quad v = - \frac{\sin(\tau)}{\cosh(\gamma) - \cos(\tau)},
\end{equation}
and the sign from Eq.~\eqref{equaphase} has been inverted, since the correct quantum phase $\mathcal{S}_\alpha(x,\tau)$ actually corresponds to the complex conjugate of Eq.~\eqref{equagaussiansuperposition}. 

One should notice the last term on the right-hand side of Eq.~(\ref{equaphase}), which corresponds to the harmonic oscillator ground state contribution. The velocity field is then written as
\begin{equation}\label{equavelfield}
\text{v}= \frac{1}{2}\frac{\partial \mathcal{S}_\alpha}{\partial x} = - \frac{vx}{2},
\end{equation}
which is $\alpha$ independent, apart from the $\gamma$-parameterization. Quantum trajectories readily follow from $d x_i(\tau)/d \tau = \text{v} (x_i(\tau))$, resulting into
\begin{equation} \label{equatrajetoriabohm}
x_i(\tau) = x_{_{0i}}\sqrt{\frac{\cosh(\gamma) -\cos(\tau) }{\cosh(\gamma) - 1}}\quad,
\end{equation}
where $ x_{_{0i}}$ corresponds to the initial position of the $i$-th particle at $\tau = 0 $. To set the correspondence with the classical trajectories, one recovers the classical dynamics from Eqs.~(\ref{equaqua42})-(\ref{equaqua44}) (for $\gamma$ given by (\ref{equagamma})) such that the classical trajectory with energy $\varepsilon$ (and $\vartheta = 0$) is retrieved for $x_{_{0i}} ^2 = \alpha + \varepsilon - \Delta$. Of course, a different classical solution would have been obtained had one considered $\varphi \neq 0$ in Eq.~\eqref{equaeqaa}, which means simply that each {\em quasi}-gaussian wave packet (for each $\varphi$) follows a different classical trajectory parameterized by $\vartheta$\footnote{The quantum trajectories from Eq.~\eqref{equatrajetoriabohm} determine (if necessary) all classically well-defined quantities, such as force, energy, and velocity although they are secondary here.}. 

\begin{figure}[t] 
\centering
\vspace{-2.cm}
\includegraphics[width=1.\linewidth]{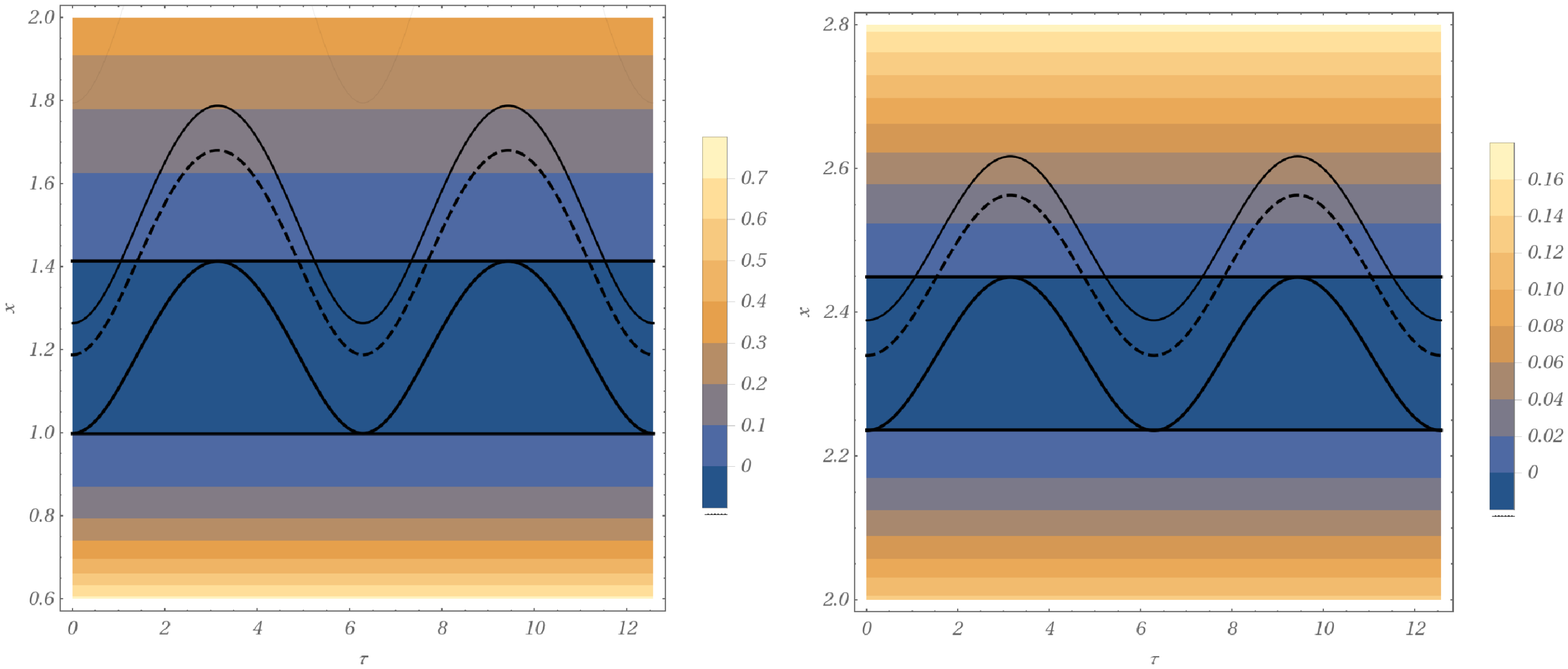}
\vspace{-3.5cm}
\caption{(Color online) Quantum trajectories superposed to the contours of the potential $\mathcal{U}(x)$. Horizontal black lines delimit the classically allowed region for $\varepsilon = 0$. From left to right, $\alpha = 3/2$ and $\alpha = 11/2$. }\label{equafigure4}
\end{figure}

The Bohmian interpretation sets that, according to the continuity equation, each trajectory determines the path along with the corresponding probability is transported. Therefore, Bohmian mechanics reproduce the statistical distribution for usual operators at arbitrary times. For the SO {\em quasi}-gaussian quantum wave packet, two particular solutions are obtained from the initial conditions given by
\begin{equation} \label{equawpcenter}
x_{_{0i}}= {u(0)}^{-1/2} \left(\alpha + 1/2 \right)^{1/2} \quad \text{and} \quad x_{_{0i}}=  {u(0)}^{-1/2}\frac{\Gamma(3/2 + \alpha)}{\Gamma(1 + \alpha)},
\end{equation}
which lead, respectively, to the description of the center of the wave packet and of the expected value of the position operator, $\langle x \rangle$.
Of course, they reproduce Schr\"odinger and Wigner formalisms as they could be obtained without the quantum trajectory formalism. However, through the trajectories above obtained, one can investigate how the potential $\mathcal{Q}(x,\tau)$ affects the dynamics. From quantum Hamilton-Jacobi equation, one has
\begin{equation}\label{equaquantumpotential}
\mathcal{Q}(x,\tau) = -\frac{1}{2}\left( x^2u^2 - 2(1 + \alpha)u + \frac{4 \alpha^2 - 1}{4x^2}\right),
\end{equation}
which is also non-vanishing for the harmonic case, as recovered by $\alpha = \pm 1/2$.
Such a modifying potential, $\mathcal{Q}(x,\tau)$, violates classical energy conservation -- which is related to the quantum tunneling phenomenon -- as depicted in Fig.~\ref{equafigure4}.
The anharmonic contribution from $\mathcal{Q}(x,\tau)$ is exactly canceled out by the potential $\mathcal{U}(x)$, and thus the quantum particle can access classically forbidden regions, as for instance, the singularity at the origin $x = 0$, with finite energy given by Eq.~\eqref{equaQHJ}.

In this case, energy is neither quantized nor constant, as usually expected for a theory of particle trajectories with a time-dependent potential\footnote{Hamiltonian eigenstates are exceptions that do exhibit energy conservation but do not yield classical motion.}. Nevertheless, considering that energy conservation is a Hamiltonian consequence for the classical case, one can assume the constraint from the classical trajectory, \[ \left( \frac{(\partial_x \mathcal{S})^2}{2} + \mathcal{U} \right) \bigg \rvert_{x = x_{cl}} = \varepsilon,\] in order to verify that $\mathcal{Q}(x, \tau)\rvert_{x=x_{cl}} \neq 0$, which in some other words asserts that the energy is not conserved along a (Bohmian) classical trajectory. Therefore, energy conservation is only asymptotically satisfied in the high-energy limit, when $ \varepsilon \gg \mathcal{Q}$. In this case, Eq.~\eqref{equatrajetoriabohm} implies that
\begin{equation}\label{equaclassicalHO2}
 x_i (\tau) = C_{0i} \bigg\rvert \sin \left(\frac{\tau}{2}\right) \bigg\rvert, 
\end{equation}
which is obtained by expanding $\exp(\pm \gamma)$ for $\gamma \ll 1$ and $C_{0i} >0$. The above equation describes classical oscillators that collide elastically at the potential barrier. A similar result is obtained for $\alpha \approx \pm 1/2$, where the same expansion can be used. The local velocity from Eq.~\eqref{equavelfield} is thus given by
\begin{equation} \label{equavelocityclassical}
 \text{v} = \frac{x\sin(\tau)}{2(1-\cos(\tau))},
\end{equation} 
and its correspondent set of trajectories is 
\begin{equation}\label{equaclassicalHO}
x_i (\tau) = C_{0i} \sin\left(\frac{\tau}{2}\right),
\end{equation}
which corresponds to classical harmonic oscillators, for negligible but finite $\mathcal{Q}$. The similarities between Eqs.~\eqref{equaclassicalHO2} and \eqref{equaclassicalHO} are noticeable. However, the former solution does not have a definite velocity at the origin, and thus one can not associate it to the velocity field from Eq.~\eqref{equavelocityclassical}. In addition, one notices the violation of the non-crossing property of Bohmian trajectories since $C_{0i}$ is no longer the initial position but it is rather related to the total energy. From these solutions, one can assert that the classical motion can be retrieved as a limiting case of Bohmian trajectories.

\begin{figure}
\centering
\includegraphics[width=1.1\linewidth]{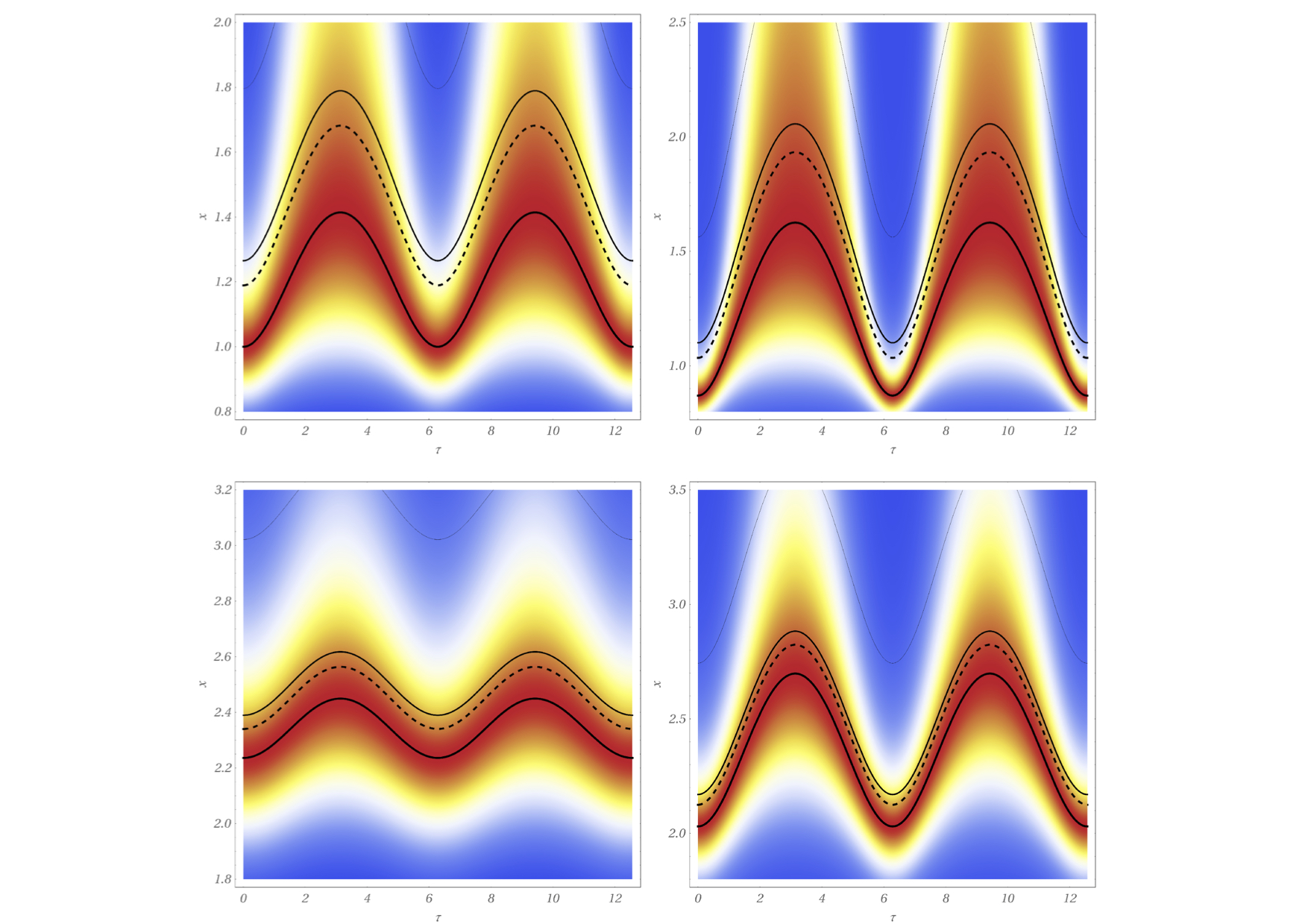}
\caption{(Color online) Bohmian trajectories for several boundary conditions: the classical solution (thickest solid line), the averaged behavior, $\langle x \rangle$ (dashed line), the wave packet center, and an arbitrary solution constrained to the tail of the distribution (thinnest solid line). The color scheme is given by 
sech($ \partial_x \mathcal{Q}$), ranging from $\partial_x \mathcal{Q} \approx 0$ (red) to unbounded values (blue). From left to right, $\varepsilon = 0$ and $\varepsilon=0.2$, and, from top to bottom, $\alpha = 3/2$ and $\alpha = 11/2$.}\label{equafigure5}
\end{figure}

Turning attention to the distortions of quantum trajectories from the classical behavior discussed above, the meaning of the particular Bohmian solutions introduced in Eq.~\eqref{equawpcenter} is cleared up as they  are superposed to $\partial_x \mathcal{Q}$ in Fig.~\ref{equafigure5}.
\noindent
For $\tau =0$, the trajectories are minimally separated and $\partial_x \mathcal{Q}$ is maximal. In this case, localized wave packets are considered to be {\em quasi}-classical, at least for short times. However, they eventually spread out under the influence of the quantum force, and Bohmian trajectories depart from classical-like behavior. The wave packet spreads out until $\tau = \pi$, when the motion is reversed, and at $\tau = 2\pi$ the initial coherence is recovered. As $\alpha$ decreases, the expected value of $\langle x \rangle$ lies in regions of intermediate values of $\partial_x \mathcal{Q}$ (white and yellow regions). Accordingly, for the opposite situation, it follows the classical trajectory. This is quantitatively verified by computing their initial separation with Eq.~\eqref{equawpcenter}, which vanishes for increasing $\alpha$.

The fact that $\langle x \rangle$ follows the classical trajectory is also observed for coherent states of the harmonic oscillator \cite{Glauber}, which do not spread over time. Here, however, the variance of the canonical variable $x$ is also a Bohmian trajectory and thus it is time dependent. As it is well known, the classical limit in this sense is far from a complete suppression of quantum phenomena --- even for the quantum harmonic oscillator --- related here to the finiteness of $\mathcal{Q}(x,\tau)$.

\begin{figure}
\centering
\includegraphics[width=0.8\linewidth]{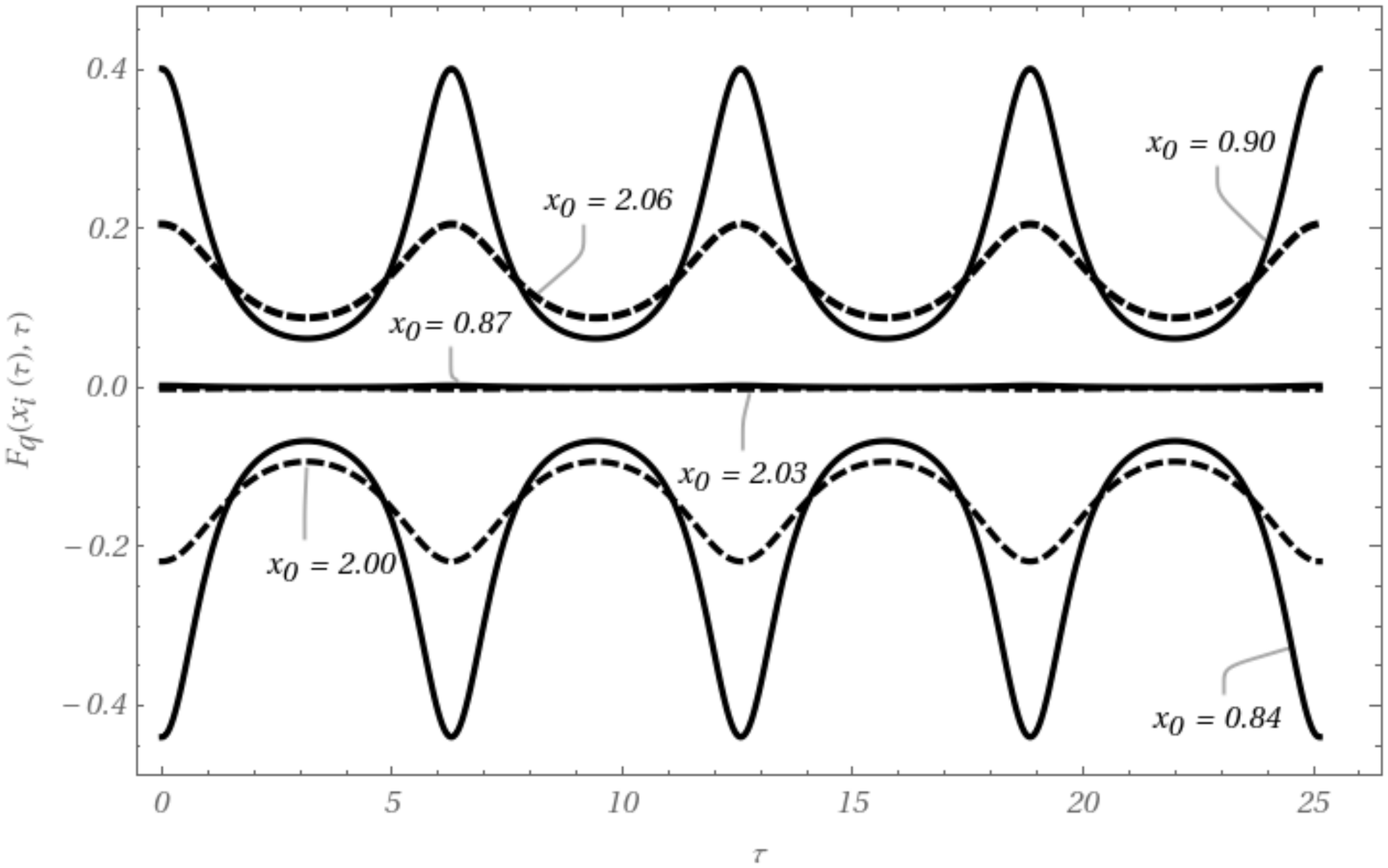}
\caption{``Quantum force'' $F_q (x_i (\tau), \tau) = - \partial_x \mathcal{Q}(x,\tau) \rvert_{x = x_i (\tau)}$ along Bohmian trajectories close to the classical solution. Solid line for $\alpha = 3/2$ and dashed line for $\alpha = 11/2$. All trajectories have the same energy parameter $\varepsilon = 0.2$ and $x_0 \equiv x_{0i}$. }\label{equafigure6}
\end{figure}

As already discussed, energy conservation is a strict condition to be imposed onto a quantum trajectory, and one can still obtain Newtonian-like motion for $\partial_x \mathcal{Q}(x_{cl} (\tau),\tau) = 0$ as it is suggested by the red region around the classical solution in Fig.~\ref{equafigure5}. Therefore, as the fine details of the quantum force are shown in Fig.~\ref{equafigure6}, 
\noindent
it confirms that $\partial_x Q$ is exactly zero along the classical solutions. To make such an assertion more clear, two equidistant solutions from the classical trajectories have been included in Fig.~\ref{equafigure6}. One notices that the anharmonic parameter $\alpha$ is associated to the stability of the wave packet: for smaller values of $\alpha$, the wave packets tend to quickly lose their {\em quasi}-gaussian property due to greater values of the quantum force, $-\partial_x \mathcal{Q}$. In this sense, Figs.~\eqref{equafigure5} and \eqref{equafigure6} are complementary: the wave packet exhibits the {\em quasi}-gaussian profile for $\tau = 2m \pi$ ($m$ integer) with maximal $\rvert \partial_x \mathcal{Q} \rvert$. 

To summarize the discussion above, quantum trajectories have a Newtonian-like dynamics for the expected harmonic case. Moreover, for increasing values of the energy parameter, the effects of the new potential term becomes less relevant and the classical case is also recovered. In the Bohmian approach, the quantum features are encoded by $\mathcal{Q}$, and it can be related to non-classical phenomena like the quantum tunneling. It has translated as an acceleration of quantum trajectories, which is not deducible from the external potential, but rather from the probability distribution itself, which prevents two Bohmian trajectories to cross. The results have shown that greater values of the anharmonic parameter $\alpha$ is associated to smaller values of such quantum force, such that the SO \textit{quasi}-gaussian wave packet retains its initial shape.
In a more general sense, such a preliminary analysis involving the effects due the anharmonic parameter into the motion of pure wave packets, ratifies the phase-space interpretation that asserts that the {\em quasi}-gaussian superposition for the SO recovers the classical scenario, reinforcing the role of such non-linear quantum system in the understanding of the classical-quantum boundaries in Nature.

\section{Maximal mixing and the thermodynamic limit}\label{equathermo}

In order to describe a quantum statistical ensemble at a finite temperature, $\mathcal{T}$,
one can map the Weyl-Wigner description onto the quantum propagator Green's function given by
\begin{equation}
\Delta(x,\tau;\,x^{\prime},0) = \langle x \vert \exp(-i\,\tau\, \hat{\mathcal{H}}) \vert x^{\prime} \rangle.
\end{equation}
Through an also dimensionless Wick rotation on 
$\tau$, $\tau \rightarrow - i \, \beta\hbar\omega$, with $\beta = 1/ k_{B} \mathcal{T}$, where $k_{B}$ is the Boltzmann constant, one has
\begin{equation}
\Delta(x,\tau;\,x^{\prime},0) \xrightarrow{Wick} \,\, \Delta^{\alpha}(x,- i \, \beta\hbar\omega;\,x^{\prime},0) = \sum_n \exp(-\beta\hbar\omega\, \varepsilon_n)\,\phi^{*\alpha}_n({x})\phi^{\alpha}_n({x^{\prime}})
= \rho^{\alpha}(x,\,x^{\prime};\,\beta),
\end{equation}
where the anharmonic index $\alpha$ has been included from Eq.~\eqref{equaqua16} and $\rho^{\alpha}(x,\,x^{\prime};\,\beta)$ is the canonical ensemble density matrix in the coordinate representation, whose non-diagonal elements are set to $x \to x+y$ and $x^{\prime} \to x - y$, so as to have the Fourier transform of the thermal density matrix identified by the thermalized phase-space probability distribution \cite{Ballentine,Hillery},
\begin{eqnarray}
\label{equaalexDimW222}
\Omega^{\alpha}(x, \, k;\,\beta) &=& \pi^{-1} \int^{+\infty}_{-\infty} \hspace{-.15cm}dy\,\exp{\left(2\, i \, k \,y\right)}\,\rho^{\alpha}(x+y,\, x-y;\,\beta),
\end{eqnarray}
for which the associated quantum operator, $\hat{\Omega}^{\alpha}$, satisfies the Bloch equation \cite{Hillery},
\begin{equation}
\frac{\partial \hat{\Omega}^{\alpha}}{\partial \beta} = - \hat{\mathcal{H}}\hat{\Omega}^{\alpha} = -\hat{\Omega}^{\alpha}\hat{\mathcal{H}},
\end{equation}
with $\hat{\Omega}^{\alpha}(\beta=0) \propto \mathbb{I}$, and from which the inception of the partition function as a trace related functional is identified by
$\mathcal{Z}\equiv\mathcal{Z}(\beta) = Tr[\exp(-\beta\hbar\omega\, \hat{\mathcal{H}})] = Tr[\hat{\Omega}^{\alpha}]$, such that more enhanced calculations related to the quantum statistical properties can be evaluated through the definitive Weyl-Wigner representation which, from Eq.~\eqref{equaalexDimW222}, is then given by
\begin{eqnarray}
\mathcal{W}_\Omega^{\alpha}(x, \, k;\,\beta) &=& (\mathcal{Z}\,\pi)^{-1} \int^{+\infty}_{-\infty} \hspace{-.15cm}dy\,\exp{\left(2\, i \, k \,y\right)}\,\rho(x+y,\, x-y;\,\beta),
\end{eqnarray}
with
\begin{equation}
\mathcal{Z}(\beta) = \int^{+\infty}_{-\infty} \hspace{-.25cm}dx\,\int^{+\infty}_{-\infty} \hspace{-.25cm}dk\,\,\Omega^{\alpha}(x, \, k;\,\beta),
\end{equation}
which leads to the Wigner distribution for the thermodynamic statistical ensemble written as
\begin{eqnarray}
\label{equaalexDimW222B}
\mathcal{W}_\Omega^{\alpha}(x, \, k;\,\beta) &=&\frac{\exp(-\beta\hbar\omega)}{\mathcal{Z}(\beta)}\sum_{n=0}^{\infty} \mathcal{W}_n^{\alpha}(x, \, k)\, \exp(-2n\,\beta\hbar\omega),
\end{eqnarray}
where $\mathcal{W}_n^{\alpha}(x, \, k)$ is written as \cite{NovoPaper}
\small\begin{eqnarray}\label{equaDimW2}
\mathcal{W}_n^{\alpha}(x, \, k) 
&=&2 (N_n^{(\alpha)})^2\, \pi^{-1} \int^{+\infty}_{-\infty} \hspace{-.15cm}dy\,\Theta(x+y)\Theta(x-y)\,(x^2-y^2)^{\frac{1}{2}+\alpha}\, \exp\left(2\,i\, k\,y\right)\\
&&\qquad\qquad\qquad\qquad\qquad\qquad\qquad\exp\left[-(x^2+y^2)\right] L_n^{\alpha}\left((x+y)^2\right)\,L_n^{\alpha}\left((x-y)^2\right)
\nonumber\\
&=& \frac{2}{\pi} \int^{+x}_{-x} \hspace{-.15cm}dy\,\exp\left(2\,i\, k\,y\right)\,\exp\left[-(x^2+y^2)\right] \,\sum_{j=0}^n \frac{L_{n-j}^{\alpha+2j}\left(2 (x^2+y^2)\right) }{\Gamma(\alpha+j+1)}
\frac{(x^2-y^2)^{\frac{1}{2}+\alpha+2j}}{j!},\nonumber
\end{eqnarray}\normalsize
since 
\begin{equation}
L_n^{\alpha}\left(x\right)\,L_n^{\alpha}\left(y\right) = \frac{\Gamma(n+\alpha+1)}{n!}\,\sum_{j=0}^n \frac{L_{n-j}^{\alpha+2j}\left(x+y\right) }{\Gamma(\alpha+j+1)}
\frac{x^j y^j}{j!}.
\end{equation}
After some simple mathematical manipulations, Eq.~\eqref{equaalexDimW222B} becomes \cite{Proceedings}
\small\begin{eqnarray}
\mathcal{W}_\Omega^{\alpha}(x, \, k;\,\beta) &=& \frac{2 \exp(-{\beta\hbar\omega})}{\mathcal{Z}(\beta)\,\pi}
\int^{+x}_{-x}
\hspace{-.15cm}dy\,\exp\left(2\,i\, k\,y\right)\,\exp\left[-(x^2+y^2)\right]\,(x^2-y^2)^{\frac{1}{2}+\alpha}\times\\
&&
\sum_{n=0}^{\infty}\bigg{\{}\exp(-2n\,\beta\hbar\omega)\frac{n!}{\Gamma(\alpha+n+1)}L_n^{\alpha}\left((x+y)^2\right)\,L_n^{\alpha}\left((x-y)^2\right)\bigg{\}},\quad\quad\quad
\nonumber
\end{eqnarray}\normalsize
with the sum in the last line being summarized by \cite{Gradshteyn}
\begin{eqnarray}
\label{equaalexDimW222C}
\frac{(x^2-y^2)^{-\alpha}}{(1-\lambda)\lambda^{\frac{\alpha}{2}}}
\exp\left[-\frac{2\lambda}{1-\lambda}(x^2+y^2)\right] \mathcal{I}_{\alpha} \left(\frac{2\lambda^{\frac{1}{2}}}{1-\lambda}(x^2-y^2)\right),\quad\quad\quad\nonumber
\nonumber
\end{eqnarray}\normalsize
where $\lambda = \exp(-{2\beta\hbar\omega})$ and $\mathcal{I}_{\alpha}$ is the {\em modified Bessel function of the first kind}. 
The above manipulations result into an integral representation of the thermalized Wigner function for the SO given by \cite{Proceedings}
\begin{eqnarray}\label{equafinalwigner}
\mathcal{W}_\Omega^{\alpha}(x, \, k;\,\beta) &=& \frac{\exp({\alpha\beta\hbar\omega})}{\sinh(\beta\hbar\omega)\mathcal{Z}(\beta)\,\pi}\int^{+x}_{-x}
\hspace{-.15cm}dy\,\exp\left(2\,i\, k\,y\right)\,(x^2-y^2)^{\frac{1}{2}}\times
\\&&
\qquad\qquad\exp\left[-\coth(\beta\hbar\omega)(x^2+y^2)\right]\,\mathcal{I}_{\alpha} \left(\frac{x^2-y^2}{\sinh(\beta\hbar\omega)}\right),\nonumber
\end{eqnarray}
which can be further simplified in the low-temperature limit (cf. Appendix III).

The partition function $\mathcal{Z}(\beta)$ follows from the phase-space integration of Eq.~\eqref{equaalexDimW222B}, which yields
\begin{eqnarray} \label{equapartition}
\mathcal{Z}(\beta) =\exp(-\,\beta\hbar\omega)\sum_{n=0}^{\infty}\exp(-2n\,\beta\hbar\omega) = \frac{\exp(-\,\beta\hbar\omega)}{1- \exp(-2\,\beta\hbar\omega)} = \frac{1}{2\sinh(\beta\hbar\omega)},
\end{eqnarray}
which corresponds to the same partition function of the harmonic oscillator with a characteristic frequency $\omega'= 2 \omega$.
For the thermalized statistical mixing described by $\mathcal{W}_\Omega^{\alpha}(x, \, k;\,\beta)$, one can also compute the (dimensionless) quantum purity through Eq. (\ref{Puritys}) (cf. Appendix II), which yields \cite{Proceedings}
\small\begin{eqnarray}\label{equafinalwigner22}
\mathcal{P}^{\alpha}(\beta) &=& \frac{8\exp({2\alpha\beta\hbar\omega})}{\pi}
\int_{0}^{\infty}\hspace{-.3 cm}dx\,
\int^{+x}_{-x}\hspace{-.15cm} dz\,
\int^{+x}_{-x}\hspace{-.15cm}dy\,
\int_{-\infty}^{+\infty}\hspace{-.3 cm}dk \exp\left(2\,i\, k\,(y+z)\right)\times\\
&&\qquad\qquad
\left[(x^2-y^2)(x^2-z^2)\right]^{\frac{1}{2}}
\exp\left[-\coth(\beta\hbar\omega)(2x^2+y^2+z^2)\right]\times\nonumber\\
&&\qquad\qquad\qquad\qquad\qquad\qquad\,\mathcal{I}_{\alpha} \left(\frac{x^2-y^2}{\sinh(\beta\hbar\omega)}\right)
\,\mathcal{I}_{\alpha} \left(\frac{x^2-z^2}{\sinh(\beta\hbar\omega)}\right)\nonumber\\
&=& 8\exp({2\alpha\beta\hbar\omega})
\int_{0}^{\infty}\hspace{-.3 cm}dx\,
\int^{+x}_{-x}\hspace{-.15cm} dz\,(x^2-z^2)\exp\left[-2\coth(\beta\hbar\omega)(x^2+z^2)\right]
\mathcal{I}^{2}_{\alpha} \left(\frac{x^2-z^2}{\sinh(\beta\hbar\omega)}\right)\nonumber\\
&=& 8\exp({2\alpha\beta\hbar\omega})
\int^{+1}_{-1}\hspace{-.15cm} ds\,(1-s^2)
\int_{0}^{\infty}\hspace{-.3 cm}dx\,x^3
\exp\left[-2x^2\coth(\beta\hbar\omega)(1+s^2)\right]
\mathcal{I}^{2}_{\alpha} \left(x^2\frac{1-s^2}{\sinh(\beta\hbar\omega)}\right)\quad\nonumber
\end{eqnarray}\normalsize
which, for half-integer values of $\alpha$, results into
\small\begin{eqnarray}\label{equafinalwigner2}
\mathcal{P}^{\alpha}(\beta) &=&\frac{1}{2^{2\alpha-1}\sqrt{\pi}}\frac{\Gamma(\alpha+3/2)}{\Gamma(\alpha+1)}
\exp({2\alpha\beta\hbar\omega})\,\tanh^{2}(\beta\hbar\omega)\,\mbox{sech}^{2\alpha}(\beta\hbar\omega)
\times\\
&&\qquad\qquad
\int^{+1}_{-1}\hspace{-.15cm} ds\,\frac{(1 - s^2)^{2\alpha+1}}{(1 + s^2)^{2\alpha+2}}\,
_2\mathcal{F}_1\left[\alpha+1/2,\,\alpha+3/2,\,2\alpha+1,\,\left(\frac{1 - s^2}{1 + s^2}\mbox{sech}(\beta\hbar\omega)\right)^2
\right].\nonumber
\end{eqnarray}\normalsize
Once introducing the series representation of the {\em ordinary hypergeometric function} $_2\mathcal{F}_1[...]$ \cite{Gradshteyn}, 
\footnotesize
\begin{equation} \label{equaseries}
\sum_{k=0} ^{+\infty} \frac{\Gamma(\alpha + k + 1/2)\, \Gamma(\alpha + k + 3/2) \, \Gamma(2\alpha + 1)}{\Gamma(\alpha + 1/2)\, \Gamma(\alpha + 3/2)\,\Gamma(2\alpha + 1 + k)\Gamma(k+1)} \left(\frac{1 - s^2}{1 + s^2}\mbox{sech}(\beta\hbar\omega)\right)^{2k},
\end{equation}
\normalsize
the integral over $s$ can be recast in the form of
\small\begin{eqnarray}\label{equacf1}
\int^{+1}_{-1}\hspace{-.15cm} ds\,\frac{(1-s^2)^{2\alpha + 2k + 1}}{(1+s^2)^{2\alpha + 2k + 2}} &=& \sqrt{\pi} \frac{\Gamma( 2 + 2k + 2\alpha)}{\Gamma(5/2 + 2k + 2\alpha)}\, _2\mathcal{F}_1\bigg[1/2,\,2 + 2k + 2\alpha,\,5/2 + 2k + 2\alpha,\,-1 \bigg]\nonumber \\
&=& \frac{\sqrt{\pi} \,\Gamma(1 + \alpha + k)}{2 \, \Gamma(3/2 + \alpha + k)},
\end{eqnarray}
\normalsize
where the second equality follows from Kummer's theorem. Inserting the above result into Eq.~\eqref{equaseries} and noticing that \[\Gamma(\alpha + k + 1) = \frac{\Gamma(2\alpha + 2k + 1)\sqrt{\pi} }{2^{2\alpha + 2k}\Gamma(\alpha + k + 1/2)},\] again, for a half-integer $\alpha$, the remaining $k$-dependent terms can be re-summed by
\small\begin{eqnarray}\label{equacf2}
\sum_{k=0} ^{+\infty} \frac{z^k}{\Gamma(k+1)\, 2^{2k}} \frac{\Gamma( 2\alpha + 2k + 1)}{\Gamma(2\alpha + k + 1)} &=& _2\mathcal{F}_1\bigg[\alpha + 1/2,\,\alpha + 1,\,2\alpha + 1,\,z \bigg]\bigg{\rvert}_{z=\text{sech}^2 (\alpha \beta \hbar \omega)} \nonumber \\
&=& \frac{2 ^{2 \alpha} \,\left(1 + \tanh(\beta \hbar \omega)\right)^{-2\alpha}}{\tanh(\beta \hbar \omega)},
\end{eqnarray}
\normalsize
where, in the last step, one has noticed that \cite{hypergeo}
\begin{equation}
_2\mathcal{F}_1\bigg[\alpha + 1/2,\,\alpha + 1,\,2\alpha + 1,\, 4y(1-y) \bigg] = \frac{1}{(1-y)^{2\alpha}(1 - 2y)}.
\end{equation}
Finally, by replacing the results from Eqs.~\eqref{equaseries}-\eqref{equacf2} into Eq.~\eqref{equafinalwigner2}, the expression for quantum purity returns $\mathcal{P}^\alpha(\beta) \equiv \mathcal{P}(\beta) = \tanh(\beta \hbar \omega)$, which is also independent of the interaction parameter $\alpha$, and is constrained to the interval between $0$ (maximal mixing) and $1$ (pure state), as expected.

\subsection{Thermalized Wigner currents}

The fluid analog of the Wigner flow is evinced when it is cast into the phase-space equivalent form of the Schr\"{o}dinger equation, which corresponds to the continuity equation \cite{NovoPaper,Steuernagel3} written as (cf. Appendix I),
\begin{equation} \label{equacontinuity}
\frac{\partial \mathcal{W}}{\partial \tau} + \frac{\partial \mathcal{J}_x}{\partial x}+\frac{\partial \mathcal{J}_k}{\partial k} = \frac{\partial \mathcal{W}}{\partial \tau} + \mbox{\boldmath $\nabla$}_{\xi}\cdot\mbox{\boldmath $\mathcal{J}$} =0,
\end{equation}
for the dimensionless current components given by
\small\begin{eqnarray}
\label{equaalexDimWA}\mathcal{J}_x(x, \, k;\,\tau) &=& k\,\mathcal{W}(x, \, k;\,\tau)
,\\
\label{equaalexDimWB}\mathcal{J}_k(x, \, k;\,\tau) &=& -\sum_{\eta=0}^{\infty} \left(\frac{i}{2}\right)^{2\eta}\frac{1}{(2\eta+1)!} \, \left[\left(\frac{\partial~}{\partial x}\right)^{2\eta+1}\hspace{-.5cm}\mathcal{U}(x)\right]\,\left(\frac{\partial~}{\partial k}\right)^{2\eta}\hspace{-.3cm}\mathcal{W}(x, \, k;\,\tau).
\end{eqnarray}\normalsize
Considering that, for the phase-space coordinate vector, $\boldsymbol{\xi} = (x,k)$, one straightforwardly identifies the corresponding classical velocity vector obtained from Hamilton's equations as $\mbox{\textbf{v}}_\xi = (k, - \partial \mathcal{U} / \partial x)$, one notices that classical phase-space distributions follow the Liouville continuity equation if all contributions for $\eta > 0$  at (\ref{equaalexDimWB}) vanish \cite{Steuernagel3,Liouvillian,EPL18,JCAP18,NovoPaper,NovoPaper02}.

The thermalized SO Wigner currents follow readily by replacing the integral representation of the Wigner function (Eq.~\eqref{equafinalwigner}) into Eqs.~\eqref{equaalexDimWA} and \eqref{equaalexDimWB}. For the $k$-component, one notices that momentum derivatives are written as
\begin{eqnarray}\label{equacurrent1}
\left(\frac{\partial }{\partial k}\right)^{2 \eta} \mathcal{W}^\alpha _{\Omega} (x,k) &=& \frac{2 \exp(\alpha \beta \omega \hbar)}{\pi} \int ^x _{-x} \,dy \, (2 i y)^{2 \eta} \exp(2 i k y) (x^2 -y^2)^{1/2}  \times \nonumber \\ &&\quad\quad\quad\quad\quad \exp\left[-\cosh(\beta \omega \hbar)(x^2 +y^2)\right] I_\alpha \left(\frac{x^2 -y^2}{\sinh( \beta \omega \hbar)}\right),
\end{eqnarray}
and, preliminarily considering the $1/x^2$ contribution from $\mathcal{U}(x)$, one has the following identities,
\begin{equation}
\left(\frac{\partial }{\partial x}\right)^{2 \eta + 1} \left(\frac{1}{x^2} \right) = -(2\eta+2) \frac{(2\eta + 1)! }{x^{2\eta + 3}},
\end{equation}
\begin{eqnarray}
\sum ^\infty _{\eta = 0} \left( \frac{i}{2}\right)^{2 \eta} \, (2 i y)^{2 \eta} \frac{2 \eta + 2}{x^{2\eta + 3}} &=& \frac{1}{x^3} \sum ^\infty _{\eta = 0} (2 \eta + 2)\epsilon^{\eta}, \quad \text{for} \quad \epsilon = \left(\frac{y}{x}\right)^2, \nonumber \\
 &=& \frac{2}{x^3} \frac{d}{d \epsilon} \sum ^\infty _{\eta = 0} \epsilon^{\eta + 1} = \frac{2}{x^3} \frac{d}{d \epsilon} \left[\frac{\epsilon}{1 - \epsilon}\right] = \frac{2}{x^3} \frac{1}{(1 - \epsilon)^2}.
\end{eqnarray}
By plugging back Eq.~\eqref{equacurrent1} and the above result with $\epsilon = (y/x)^2$ into the Wigner current expression from Eq.~\eqref{equaalexDimWB}, it returns
\small
\begin{equation}
\frac{4 \, x \exp(\alpha \beta \hbar \omega)}{\pi} \int ^x _{-x} dy \frac{\exp(2 i k y)\,(x^2 - y ^2)^{1/2} }{(x^2 - y ^2)^2} \exp\left[-\coth(\beta \hbar \omega)(x^2 + y ^2)\right]I_\alpha \left(\frac{x^2 -y^2}{\sinh( \beta \omega \hbar)} \right).
\end{equation}
\normalsize
Now, considering the following iterative property for $I_\alpha (z)$,
\begin{equation}
\frac{I_\alpha (z)}{z} = \frac{1}{2\alpha} \bigg(I _{\alpha - 1} (z) - I_{\alpha + 1} (z) \bigg),
\end{equation}
which can be used twice to obtain $I_\alpha (z) / z^2$, one has
\footnotesize
\begin{eqnarray}\label{equafinalcurrent}
 &=& \frac{4 \, x \exp(\alpha \beta \hbar \omega)}{\pi\sinh ^2 ( \beta \omega \hbar)} \frac{1}{4\alpha(\alpha - 1)(\alpha +1)}\int ^x _{-x} dy \,\exp(2 i k y)\,(x^2 - y ^2)^{1/2} \exp\left[-\coth(\beta \hbar \omega)(x^2 + y ^2)\right] \, \times \nonumber \\
&& \qquad (\alpha + 1)\,I _{\alpha - 2} \left(\frac{x^2 - y ^2}{\sinh( \beta \omega \hbar)}\right) - 2 \alpha \, I_{\alpha} \left(\frac{x^2 - y ^2}{\sinh( \beta \omega \hbar)}\right) + (\alpha - 1)\, I _{\alpha + 2} \left(\frac{x^2 - y ^2}{\sinh( \beta \omega \hbar)}\right) \nonumber \\ 
&=& \frac{-x \mathcal{W}_{\Omega} ^\alpha}{\sinh ^2 ( \beta \omega \hbar)(\alpha - 1)(\alpha + 1)} + \frac{ x \mathcal{W}_{\Omega} ^{\alpha-2} \exp( 2 \beta \hbar \omega) }{2 \sinh ^2 ( \beta \omega \hbar) \alpha (\alpha - 1)} + \frac{ x \mathcal{W}_{\Omega} ^{\alpha+ 2} \exp(- 2 \beta \hbar \omega) }{2 \sinh ^2 ( \beta \omega \hbar) \alpha (\alpha + 1)}.
\end{eqnarray}
\normalsize
which, once multiplied by $(4\alpha^2 - 1)/8$, is added to the harmonic contribution so as to yield the complete expression for the Wigner current components,
\begin{equation}\label{equacurrentfinala}
\mathcal{J}_x ^\alpha = k \mathcal{W}_{\Omega} ^\alpha 
\end{equation}
and
\begin{eqnarray}\label{equacurrentfinalb}
\mathcal{J}_k ^\alpha &=& - x \mathcal{W}_{\Omega} ^\alpha + \frac{4\alpha^2 - 1}{8}
\left( - \frac{x \mathcal{W}_{\Omega} ^\alpha}{\sinh ^2 ( \beta \omega \hbar)(\alpha - 1)(\alpha + 1)} \right. \nonumber\\
 &&\left. \qquad\qquad\qquad\qquad\qquad + \frac{ x \mathcal{W}_{\Omega} ^{\alpha-2} \exp( 2 \beta \hbar \omega) }{2 \sinh ^2 ( \beta \omega \hbar) \alpha (\alpha - 1)}+\frac{ x \mathcal{W}_{\Omega} ^{\alpha+ 2} \exp(- 2 \beta \hbar \omega) }{2 \sinh ^2 ( \beta \omega \hbar) \alpha (\alpha + 1)} \right).
\end{eqnarray}
\begin{figure}[t] 
\centering
\includegraphics[width=0.78\linewidth]{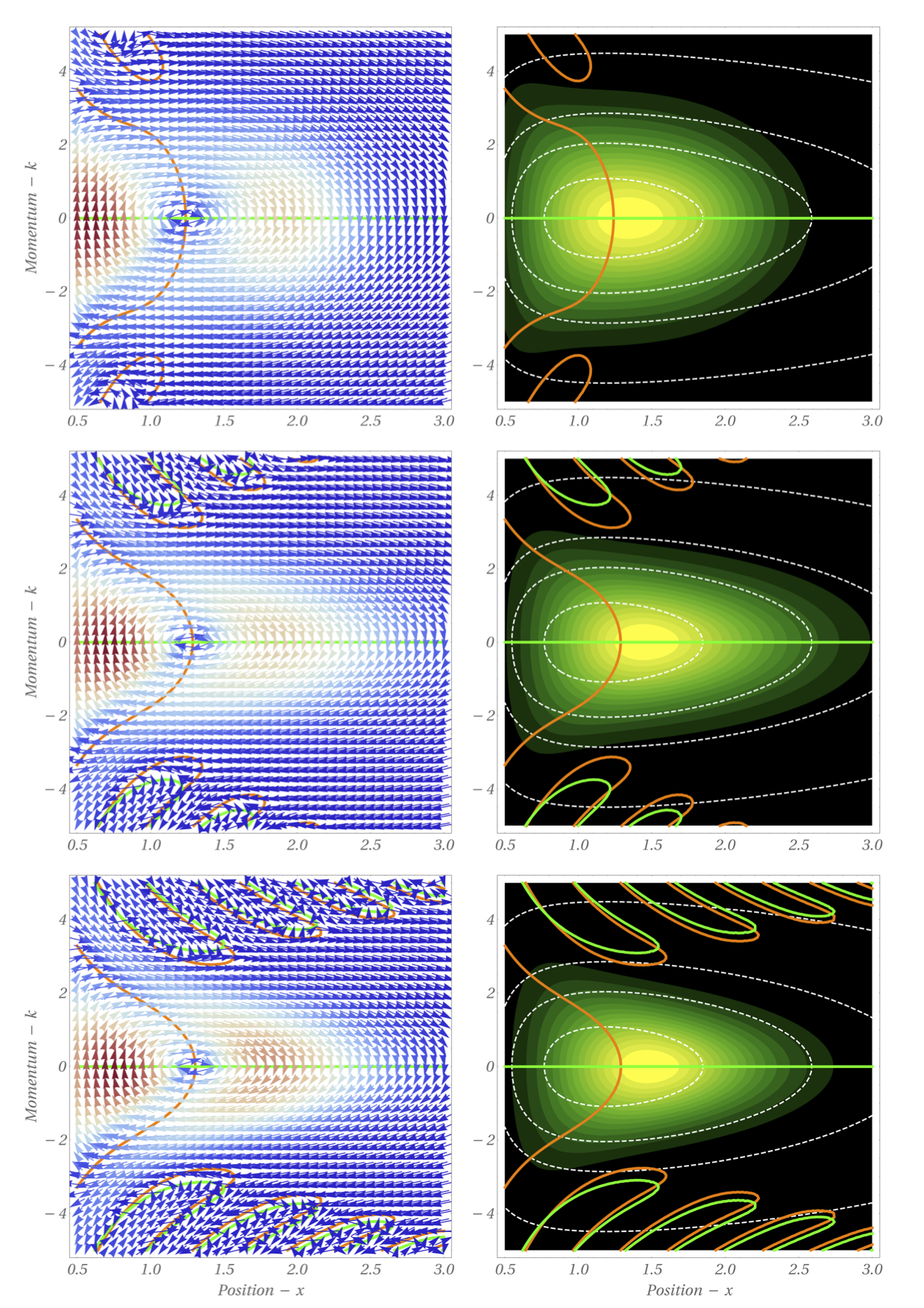}
\caption{(Color online) Wigner flow in the phase-space. (Left column) The color scheme describes the normalized modulus of the Wigner current $\mbox{\boldmath $\mathcal{J}$}^\alpha$ from 0 (blue) to 1 (red). (Right column) Contour lines corresponding to regions of flow reversal in the $k$ direction for $\mathcal{J}_k = 0$ (orange) and in the $x$ direction for $\mathcal{J}_x = 0$ (green), respectively. The background is superposed by classical trajectories (white dashed lines) and the results are all for $\alpha = 3/2$. Orange and green line intersections correspond to stagnation points of the Wigner flow. From top to bottom, $\beta = 0.5 (\hbar \omega)^{-1}, (\hbar \omega)^{-1}, 2(\hbar \omega)^{-1}$.}\label{equafigure1}
\end{figure}
\begin{figure}[t] 
\centering
\vspace{-2.cm}
\includegraphics[width=1.\linewidth]{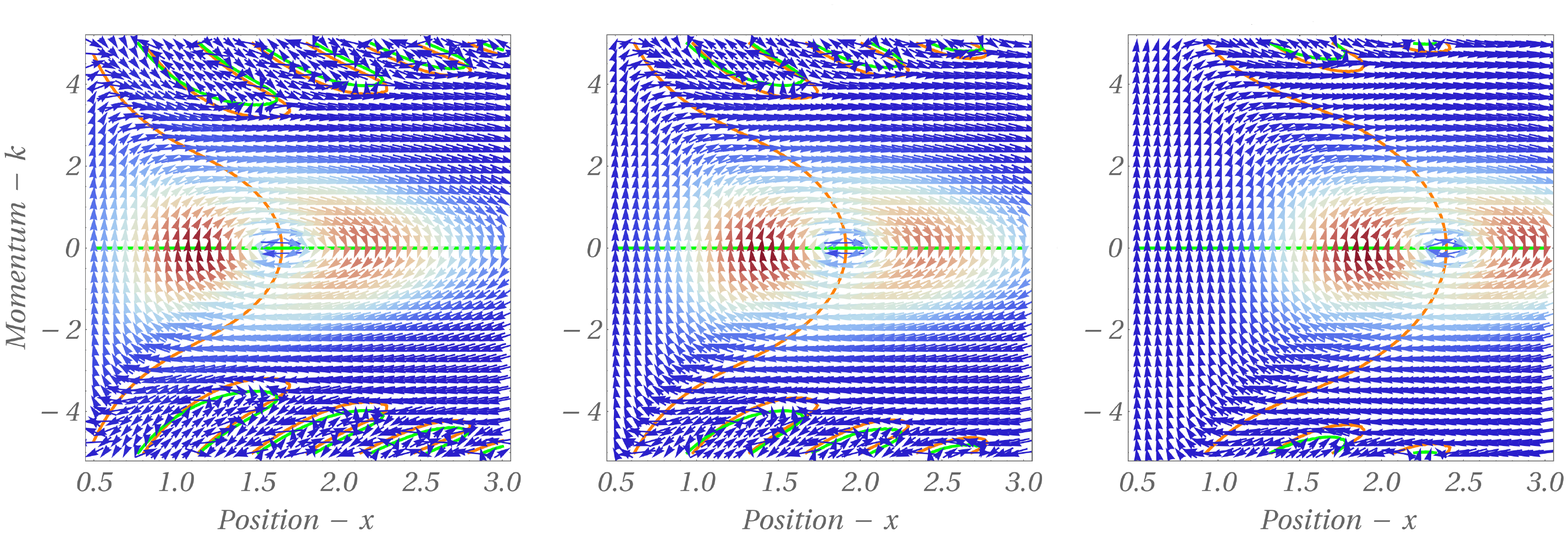}
\vspace{-3.5cm}
\caption{(Color online) Qualitative influence of the anharmonic parameter, $\alpha$, onto phase-space flow. The color scheme is the same used in Fig.~\ref{equafigure1}, from 0 (blue) to 1 (red) for the modulus of $\mbox{\boldmath $\mathcal{J}$} ^\alpha$. Orange and green lines correspond to $\mathcal{J}_k = 0$ and $\mathcal{J}_x = 0$, respectively. The temperature is held constant with $\beta = 2 (\hbar \omega)^{-1}$ and, from left to right, $\alpha = 5/2,\, 7/2,\, 11/2$.}\label{equafigure2}
\end{figure}
The classical Wigner current is similarly obtained by truncating the series expansion in Eq.~\eqref{equaalexDimWB} at the first term\footnote{which is simply $\mbox{\boldmath $\mathcal{J}$} ^{\alpha (cl)} = \left(k, - \frac{\partial \mathcal{U}}{\partial x}\right) \mathcal{W}^\alpha =\left(k, -x -\frac{1-4\alpha^2}{4x^3}\right)\mathcal{W}^\alpha$.}. Therefore, higher order derivatives of the anharmonic potential mix contributions from different quantum mixtures parameterized by $\alpha$. This induces additional flow stagnation points even for non-vanishing values of the Wigner function. Also, the Wigner current is temperature dependent, which is not observed in the classical current apart from the temperature dependence of the Wigner function itself. Such behavior is depicted in Figs.~\eqref{equafigure1} and \eqref{equafigure2}, where stagnation points are identified for orange-green crossing lines for $\mathcal{J}_k ^\alpha = \mathcal{J}_x ^\alpha = 0$. Therefore, the evolution of the quantum flow is counterbalanced by the orange and green fringes, which delimit the regions of flow reversal. Otherwise, the Liouvillian flow is retrieved in the harmonic case, for which $\mathcal{W}^{\, \alpha = \pm 1/2}(x,k;\beta)=0$ always implies coincident orange and green lines for $\mathcal{J}_x = \mathcal{J}_k = 0$ instead of isolated points \cite{Steuernagel3}. There is a small caveat in that $\alpha = \pm 1/2$ does not retrieve the classical probability distribution, even though the classical and quantum phase-space distributions follow the Liouville equation. Indeed, even in the harmonic case, the Wigner function depends on the quantum state and therefore on the mixing parameter. Thus, it can still assume negative values, in a kind of non-classical feature. Apart from the quantum nature of the \textit{quasi}-distribution itself, greater values of the anharmonic parameter brings the system to the classical regime (cf. Fig.~\ref{equafigure2}) -- something similar to what has been verified for pure states \cite{NovoPaper,JCAP18}. However, quantum fluctuations have a $\beta$-dependent amplitude, which vanishes for increasing values of the temperature (cf. Fig.~\ref{equafigure1}).

As already pointed out, the classical flow vector $\mbox{\boldmath $\mathcal{J}$}^{(cl)} = \mbox{\textbf{v}}_\xi \mathcal{W}$ implies a Hamiltonian system, for which Eq.~\eqref{equacontinuity} reduces to the classical Liouville equation, since $\mbox{\boldmath $\nabla$}_\xi \cdot \mbox{\textbf{v}}_\xi = 0 $. Similarly, the quantum flow vector can be  written as $\mbox{\boldmath $\mathcal{J}$} = \mbox{\textbf{w}} \mathcal{W}$. Through this approach, the quantum phase-space velocity, $\mbox{\textbf{w}}$, encompasses the distortions over a Liouvillian flow. In particular, the right side of Fig.~\ref{equafigure1} evinces the departure of the Wigner flow from classical trajectories, illustrated by the white dashed lines. Thus, for a quantum velocity $\mbox{\textbf{w}}$, which describes generally a non-Liouvillian fluid, its divergence is written in terms of the quantum flux vector \cite{EPL18,JCAP18} as
\begin{equation}\label{equaalexquaz59}
\mbox{\boldmath $\nabla$}_\xi \cdot \mbox{\textbf{w}}= \frac{\mathcal{W} \mbox{\boldmath $\nabla$}_\xi \cdot \mbox{\boldmath $\mathcal{J}$} - \mbox{\boldmath $\mathcal{J}$} \cdot \mbox{\boldmath $\nabla$}_\xi \mathcal{W} }{\mathcal{W}^2},
\end{equation}
which, of course, is non-vanishing in the quantum (non-linear) regimes. Therefore, $\mbox{\boldmath $\nabla$}_\xi \cdot \mbox{\textbf{w}}$ is suitable for locally detecting and quantifying non-Liouvillian flows. 

\begin{figure}[t] 
\centering
\vspace{-2.5cm}
\includegraphics[width=1\linewidth]{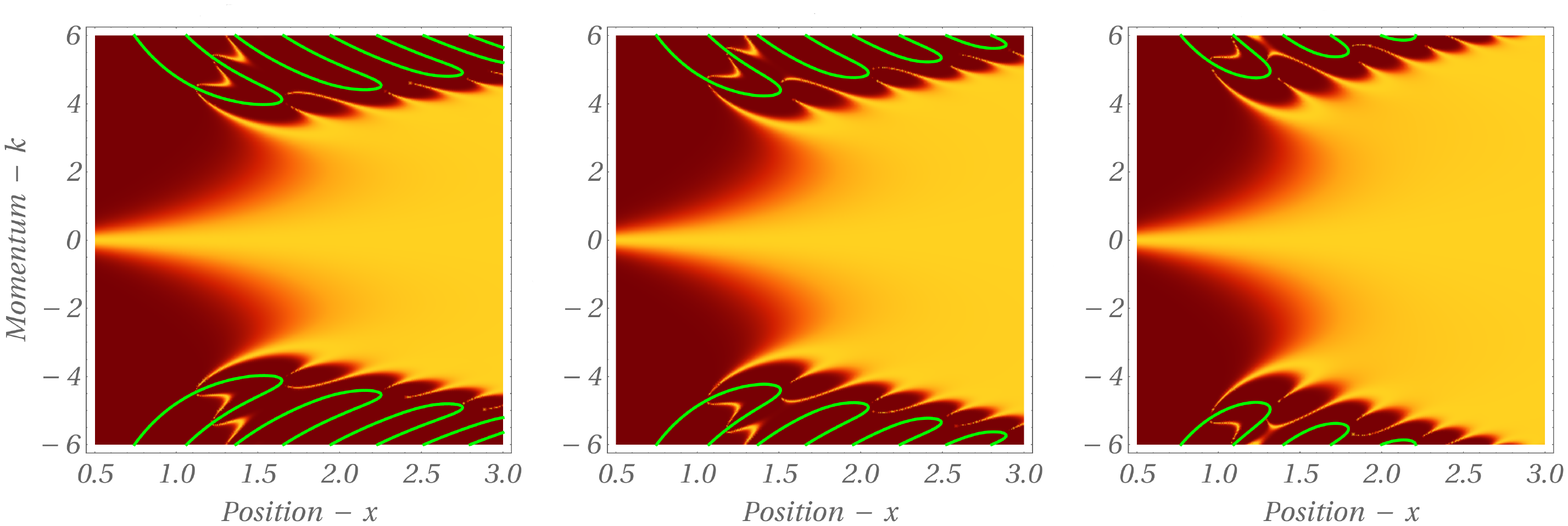}
\vspace{-3.5cm}
\caption{(Color online) Liouvillian behavior as function of the temperature, $\mathcal{T}$, parameterized by 
sech$(\mbox{\boldmath $\nabla$}_\xi \cdot \mbox{\textbf{w}})$. The light-dark color scheme identifies approximately Liouvillian flows for $\mbox{\boldmath $\nabla$}_\xi \cdot \mbox{\textbf{w}} \approx 0$ (light color). The thermalized system is maximally far from Liouvillian behavior for green lines, i.e., for the contours where $\mathcal{W}(x,k;\beta) = 0$, such that $\mbox{\boldmath $\nabla$}_\xi \cdot \mbox{\textbf{w}}$ is unbounded. From left to right, $\beta = 2 (\hbar \omega)^{-1}$, $1.5(\hbar \omega)^{-1}$, $(\hbar \omega)^{-1}$ and it has been arbitrarily set $\alpha = 7/2$.}.\label{equafigure3}
\end{figure}

For the above quantum system, an analytic expression for the Liouvillian behavior quantifier $\nabla _ \xi . {\textbf w}$ for the Wigner function at thermal equilibrium can be computed from the currents obtained through Eqs.~\eqref{equacurrentfinala} and \eqref{equacurrentfinalb},
\begin{eqnarray}
\nabla _ \xi . {\textbf w} &=& \left( \frac{4 \alpha^2 -1}{8} \right) \left(\frac{ x \exp( 2 \beta \hbar \omega) }{2 \sinh ^2 ( \beta \omega \hbar) \alpha (\alpha - 1)} \frac{\partial}{\partial k} \left(\frac{\mathcal{W} ^{\alpha-2}}{\mathcal{W} ^{\alpha}}\right) \right.\nonumber\\&&\left. \qquad\qquad\qquad\qquad\qquad\qquad+ \frac{ x \exp(- 2 \beta \hbar \omega) }{2 \sinh ^2 ( \beta \omega \hbar) \alpha (\alpha + 1)} \frac{\partial}{\partial k} \left(\frac{\mathcal{W} ^{\alpha+2}}{\mathcal{W} ^{\alpha}}\right) \right),
\end{eqnarray}
which is shown in Fig.~\ref{equafigure3} from which one can notices that quantum flow becomes locally Liouvillian for increasing values of the temperature (decreasing values of $\beta$). This accounts for the suppression of the quantum back reaction effects from $\mathcal{W}^{\alpha -2}$ and $\mathcal{W}^{\alpha + 2}$ contributions.
The results presented in Figs.~\eqref{equafigure1}, \eqref{equafigure2}, and \eqref{equafigure3} are all consistent with the quantum purity analytically calculated as $\mathcal{P} = \tanh(\beta \hbar \omega)$: in the ideal limit of the pure-state case ($\mathcal{T}=0$), the Wigner profile is non-Liouvillian. For $\mathcal{T}\rightarrow \infty$, however, $\mbox{\boldmath $\nabla$}_\xi \cdot \mbox{\textbf{w}}$ goes to 0, and thermal fluctuations balance out quantum fluctuations with the suppression of stagnation points.

Non-classical aspects are locally introduced according to Eq.~\eqref{equaalexquaz59}; however, one may further investigate how they globally affect the classical flow. For this, one recovers the results from the {\em substantial derivative} integral theorem \cite{Gradshteyn,EPL18,JCAP18,NovoPaper,NovoPaper02}, from which, 
for an infinitesimal phase-space volume element identified by $dV = dx \, dk$, a periodic quantum system 
with the phase-space integrated probability flux written as
\begin{equation}
\varsigma_{(\mathcal{C})} =\int_{V_{_{\mathcal{C}}}}dV\,\mathcal{W},
\label{equaalexquaz60}
\end{equation}
where $V \rightarrow V_\mathcal{C}$, is the (bidimensional) volume enclosed by the classical path $\mathcal{C}$, 
has its time-evolution equation expressed by \cite{EPL18,JCAP18,NovoPaper,NovoPaper02}
\begin{equation}
 \frac{D\varsigma_{_{(\mathcal{C})}} }{D\tau} = \int_{V_{_{\mathcal{C}}}}dV \,\left[\mbox{\boldmath $\nabla$}_{\xi}\cdot (\mathbf{v}_{\xi}\mathcal{W}) - \mbox{\boldmath $\nabla$}_{\xi}\cdot \mbox{\boldmath$\mathcal{J}$}\right],
\label{equaalexquaz51CC}
\end{equation}
Of course, $\mathbf{v}_{\xi} \equiv \mathbf{v}_{\xi (\mathcal{C})}$ and the classical path implies an energy parameterization. To further simplify the above expression, one identifies $\Delta\mbox{\boldmath$\mathcal{J}$} = \mbox{\boldmath$\mathcal{J}$} - \mathbf{v}_{\xi}\mathcal{W}$ as the quantum correction terms in the Wigner currents, and Eq.~\eqref{equaalexquaz51CC} becomes  \cite{EPL18,JCAP18,NovoPaper,NovoPaper02} 
\begin{eqnarray}
\frac{D\varsigma_{_{(\mathcal{C})}} }{D\tau} = -\int_{V_{_{\mathcal{C}}}}dV\, \mbox{\boldmath $\nabla$}_{\xi}\cdot \Delta \mbox{\boldmath$\mathcal{J}$} &=& -\oint_{\mathcal{C}}d\ell\, \Delta\mbox{\boldmath$\mathcal{J}$}\cdot \mathbf{n} \nonumber \\ 
&=& \int_{0}^{T}d\tau\, \Delta \mathcal{J}_k(x_{_{\mathcal{C}}}\bb{\tau},\,k_{_{\mathcal{C}}}\bb{\tau};\tau)\,\,\frac{d}{d\tau}{x}_{_{\mathcal{C}}}\bb{\tau},
\label{equaalexquaz51DD}
\end{eqnarray}
where $T$ is the classical period and $\mathbf{n}$ a unit vector given by $\mathbf{n}= (-d{k}_{_{\mathcal{C}}}/d\tau, d{x}_{_{\mathcal{C}}}/d\tau) \vert\mathbf{v}_{\xi}\vert^{-1}$ and satisfying $\mathbf{n}\cdot\mathbf{v}_{\xi}= 0$. In the classical limit $\Delta\mbox{\boldmath$\mathcal{J}$}$ vanishes; therefore, Eq.~\eqref{equaalexquaz51DD} detects global non-classical phase-space profiles, which are associated to loss and gain of coherence encoded by the Wigner function through a classical trajectory. For this reason, given a classical domain, this tool systematically quantifies topological fluctuations of the Wigner flow.

From Eq.~\eqref{equafinalwigner} one notices that the thermalized Wigner function does not depend explicitly on time, and thus $\mathcal{W}^\alpha _\Omega (x_\mathcal{C}(\tau),k_\mathcal{C}(\tau);\beta)$, the Wigner function whose arguments follow Eqs.~\eqref{equaqua43} and \eqref{equaqua44}, is also a periodic function with period $2\pi$. From Eq.~\eqref{equaalexquaz51DD} , it follows that 
\begin{equation}
 \frac{D\varsigma_{_{(\mathcal{C})}}}{D\tau} = -\int_{0} ^{2\pi} d\tau\, \Delta \mbox{\boldmath$ \mathcal{J} $}^\alpha _\Omega (x_\mathcal{C}(\tau),k_\mathcal{C}(\tau);\beta)\,k_\mathcal{C}(\tau) = 0,
\end{equation}
since the integrand is an odd function on $\tau$. The conservation of probability for non-vanishing integrand along $\mathcal{C}$ is due to the parity of the Wigner currents, i.e., $\mathcal{J}_x (x,-k) = -\mathcal{J}_x (x,+k)$ and $\Delta \mathcal{J}_k (x,-k) = \Delta \mathcal{J}_k (x,k)$, which means that the stagnation points always occur in pairs and have opposite winding numbers below and above the $x$ axis. Exceptionally, Figs.~\eqref{equafigure1} and \eqref{equafigure2} show an additional stagnation point at the $x$ axis. It is formed by a clockwise vortex also expected classically, since it corresponds to the intersection $\mathcal{J}_x ^{(cl)}(x,k) = \mathcal{J}_k ^{(cl)}(x,k)=0$ given by $\partial \mathcal{U}/ \partial x = 0$ and $k=0$, i.e., a force-free region with vanishing momentum \cite{Steuernagel3}. In this case, the net effect of stagnation points average out to zero inside an arbitrary classical domain. Furthermore, similar continuity equations for purity and entropy \cite{NovoPaper, JCAP18, EPL18} would confirm that the quantum fluctuations identified here have a vanishing global effect, and thus the thermal equilibrium quantum system produces neither entropy nor purity. 
 
\section{Conclusions}\label{equaconclusions}

The SO quantum-classical boundaries have been investigated under two different perspectives, Bohmian mechanics and the Weyl-Wigner formalism, for both of which the anharmonic contribution from the quantum mechanical potential has been associated to a non-classical dynamics. On one hand, in the former approach, quantum features are associated to the so-called quantum force, which deviates quantum trajectories from classical ones. 
Quantum trajectories have been analytically derived for a {\em quasi}-gaussian wave packet, which has evinced how the anharmonic and energy parameter drive the non-classical dynamics. The simple harmonic and high-energy cases have been retrieved, which provides the expected classical trajectories. Thus, the classical-like motion has been recovered for negligible quantum fluctuations. In quantum cosmological scenarios \cite{JCAP18,Kiefer07,Bojowald}, for instance, it may provide a set of distinct quantum scale factors subject to initial data, which can replace the classical predictions in the early universe. In the particular context of HL cosmology discussed in \cite{JCAP18}, Bohmian trajectories can be re-parameterized in terms of the quantum scale factor $a(x)$, as to provide evolution equations which can probe how the classical scenario emerges from a quantum hypothesis.

On the other hand, a not so obvious but complementary result has been obtained in the context of Wigner ensembles, for which the non-classical features are associated to a non-Liouvillian flow \cite{NovoPaper,JCAP18,EPL18,Liouvillian}. The results have preliminarily shown how quantum fluctuations affect the Liouvillian flow for pure states. The more general case concerns a statistical quantum ensemble at thermal equilibrium, for which the Wigner function and currents have been computed for the thermodynamic mixing. Given the fact that the infinite series from the Wigner current has been obtained in terms of the Wigner function from different quantum mixtures, all contributions from the anharmonic potential have been taken into account and are encoded by the thermalized quantum distribution, providing a non-perturbative definitive result. It has been shown that the quantum purity is actually independent of the anharmonic parameter, which confirms that the thermalized Wigner function described here encodes equivalent measurable information to a harmonic quantum system, even though their phase-space flow profiles are distinct. The harmonic flow does not exhibit the topological fluctuations that are characteristic of non-linear quantum fluctuation effects, even though the corresponding quantum distribution is not necessarily a classical one. Surprisingly, these disturbances to the classical flow are only detected locally, where the thermal fluctuations quickly decohere the quantum system to the classical regime, and thus entropy and purity are also (phase-space closed path) conserved. In this sense, the loss of information encoded by the SO quantum system here investigated, suggests that further investigations for out-of-equilibrium quantum systems, within the same framework, and according to the methods reported here, are welcome.
Given the analytic properties regarding the characteristic of Wigner functions, in certain sense, this formalism tackles the difficult task of understanding how a classical regime emerges from a quantum description.

\vspace{.5 cm}
{\em Acknowledgments -- The work of AEB is supported by the Brazilian Agencies FAPESP (Grant No. 2018/03960-9) and CNPq (Grant No. 301000/2019-0). The work of CFS is supported by the Brazilian Agency CAPES (Grant No. 88887.499837/2020-00).}

\section*{Appendix I - {\em Quasi}-gaussian superposition pure state}\label{equaapp2}
According to Refs.~\cite{JCAP18,NovoPaper}, departing from $\mathcal{W}^{\alpha}(x,\,k;\,\tau_{})$ from Eq.~(\ref{equaeqn27}), since
\begin{eqnarray}
\int_{-\infty}^{+\infty}\hspace{-.3 cm}dk
\,\exp\left(2\,i\,x\,k\,s \right) &=& 2\pi\,\delta(2\,x\,s) = \frac{\pi}{\vert x\vert}\delta(s),
\label{equaeqnA02}
\end{eqnarray}
the integral over $s$ returns
\begin{equation}
\frac{\pi}{\vert x\vert}\int_{-1}^{+1} ds\,\delta(s) (1-s^2)^{\frac{1}{2}+\alpha}\, \exp\left(-u \, x^2\,s^2\right) \exp\left(2i\,v\, x^2\,s\right) = \frac{\pi}{\vert x\vert}
\label{equaeqnA01B},
\end{equation}
and thus the normalization condition is satisfied,
\begin{eqnarray}
\int_{0}^{\infty}\hspace{-.3 cm}dx\int_{-\infty}^{+\infty}\hspace{-.3 cm}dk \,\mathcal{W}^{\alpha}(x,\,k;\,\tau_{}) &=& \frac{2\,u^{1+\alpha}}{\Gamma(1+\alpha)}\,\int_{0}^{\infty}\hspace{-.3 cm}dx \,x^{(1+2\alpha)}\,\exp\left(-u \, x^2\right) = 1\label{equaeqnA03}.
\end{eqnarray}
Similarly, for the dimensionless quantum purity identified by,
\begin{equation}\label{Puritys}
\mathcal{P} = 2 \pi \int^{+\infty}_{-\infty} \hspace{-.25cm}dx\,\int^{+\infty}_{-\infty} \hspace{-.25cm}dk\,\,\mathcal{W}^2,
\end{equation}
one has for the \textit{quasi}-gaussian superposition \cite{JCAP18},
\begin{eqnarray}
\lefteqn{\int_{0}^{\infty}\hspace{-.3 cm}dx\int_{-\infty}^{+\infty}\hspace{-.3 cm}dk \,\left(\mathcal{W}^{\alpha}(x,\,k;\,\tau_{})\right)^{2} =\frac{4\,u^{2(1+\alpha)}}{\pi^2\,\Gamma^2(1+\alpha)}\times}\nonumber\\
&& \,\int_{0}^{\infty}\hspace{-.3 cm}dx \,x^{4(1+\alpha)}\,\exp\left(-2\,u \, x^2\right)
\,\int_{-\infty}^{+\infty}\hspace{-.3 cm}dk
\,\exp\left(2\,i\,x\,k \,(s+r) \right)\times
\nonumber\\
&& \qquad\int_{-1}^{+1} ds\int_{-1}^{+1} dr\,[(1-r^2)(1-s^2)]^{\frac{1}{2}+\alpha}\, \exp\left(-u \, x^2\,(r^2+s^2)\right) \exp\left(2i\,v\, x^2\,(s+r)\right). \quad\,
\label{equaeqnA04}
\end{eqnarray}
The integration over $k$ gives a delta function, which is integrated over $r$ afterward,
\begin{eqnarray}
\lefteqn{\int_{0}^{\infty}\hspace{-.3 cm}dx\int_{-\infty}^{+\infty}\hspace{-.3 cm}dk \,\left(\mathcal{W}^{\alpha}(x,\,k;\,\tau_{})\right)^{2} = \frac{4\,u^{2(1+\alpha)}}{\pi^2\,\Gamma^2(1+\alpha)}\times}\nonumber\\ &&\qquad\qquad\int_{0}^{\infty}\hspace{-.3 cm}dx \,x^{(3+4\alpha)}\,\exp\left(-2\,u \, x^2\right)\,\int_{-1}^{+1} ds\,(1-s^2)^{1+2\alpha}\, \exp\left(-2\,u \, x^2\,s^2\right).
\label{equaeqnA05}
\end{eqnarray}
The integral over $x$ becomes a simple gaussian times a polynomial 
\begin{eqnarray}
\int_{0}^{\infty}\hspace{-.3 cm}dx \,x^{(3+4\alpha)}\,\exp\left(-2\,u \, x^2\,(1+s^2)\right) = \frac{1}{2^{3+2\alpha}u^{2(1+\alpha)}}
\frac{\Gamma(2(1+\alpha))}{(1+s^2)^{2(1+\alpha)}},
\label{equaeqnA06}
\end{eqnarray}
and finally Eq.~\eqref{equaeqnA05} returns the pure-state constraint
\begin{eqnarray}
\int_{0}^{\infty}\hspace{-.3 cm}dx\int_{-\infty}^{+\infty}\hspace{-.3 cm}dk \,\left(\mathcal{W}^{\alpha}(x,\,k;\,\tau_{})\right)^{2}
 &=&
\frac{1}{2^{1+2\alpha}\pi}
\frac{\Gamma(2(1+\alpha))}{\Gamma^2(1+\alpha)} \int_{-1}^{+1} ds\,\frac{(1-s^2)^{1+2\alpha}}{(1+s^2)^{2(1+\alpha)}}\nonumber\\
 &=&
\frac{1}{2^{2+2\alpha}\pi}
\frac{\Gamma(2(1+\alpha))}{\Gamma^2(1+\alpha)} \frac{\sqrt{\pi} \Gamma(1+\alpha)}{2\Gamma(3/2+\alpha)}
\nonumber\\
 &=&\frac{1}{2\pi}.
\label{equaeqnA07}
\end{eqnarray}

\section*{Appendix II - Dimensionless Wigner currents} \label{equaapp1}
The time-evolution of $W(q,p;t)$ is obtained from the divergence of a flow field \cite{Case}, 
\begin{equation}
\frac{\partial W}{\partial t} + \frac{\partial J_q}{\partial q}+\frac{\partial J_p}{\partial p} \equiv
\frac{\partial W}{\partial t} + \mbox{\boldmath $\nabla$}\cdot \mathbf{J} =0,
\label{equaalexquaz51}
\end{equation}
where $\mathbf{J}(q,p;t) = \mbox{J}_q \, \hat{q} + \mbox{J}_p \, \hat{p}$ is the vector flux whose phase-space components are given by
\begin{equation}
J_q(q,\,p;\,t)= \frac{p}{m}\,W(q,\,p;\,t), \label{equaalexquaz500BB}
\end{equation}
\begin{equation}\label{equaalexquaz500}
J_p(q,\,p;\,t) = -\sum_{\eta=0}^{\infty} \left(\frac{i\,\hbar}{2}\right)^{2\eta}\frac{1}{(2\eta+1)!} \, \left[\left(\frac{\partial~}{\partial q}\right)^{2\eta+1}\hspace{-.5cm}V(q)\right]\,\left(\frac{\partial~}{\partial p}\right)^{2\eta}\hspace{-.3cm}W(q,\,p;\,t).
\end{equation}
The classical vector flux is identified if only the term $\eta = 0$ is kept. The dimensionless Wigner currents follow from the new phase-space variables introduced in Sec.~\eqref{equasec2} and are explicitly given in terms of the dimensionful quantities \cite{NovoPaper},
\begin{eqnarray}
\mathcal{W}(x, \, k;\,\omega t) &\equiv & \hbar\, W(q,\,p;\,t) \\
\omega\, \partial_x\mathcal{J}_x (x, \, k;\,\omega t)&\equiv & \hbar\, \partial_q J_q(q,\,p;\,t) \\
\omega \,\partial_k\mathcal{J}_k (x, \, k;\,\omega t) &\equiv & \hbar \,\partial_p J_p(q,\,p;\,t),
\end{eqnarray}
where $\omega t = \tau$ is the dimensionless time, and the integration volume is re-scaled as to absorb the extra $\hbar$ in the right-hand side. Finally, Eqs.~\eqref{equaalexquaz500BB} and \eqref{equaalexquaz500} are re-written as 
\small\begin{eqnarray}\label{equaalexDimW}
\mathcal{W}(x, \, k;\,\tau) &=& \pi^{-1} \int^{+\infty}_{-\infty} \hspace{-.15cm}dy\,\exp{\left(2\, i \, k \,y\right)}\,\phi(x - y;\,\tau)\,\phi^{\ast}(x + y;\,\tau),\quad \mbox{with $y = \left(m\,\omega\,\hbar^{-1}\right)^{1/2} w$},\,\,\,\,\\
\mathcal{J}_x(x, \, k;\,\tau) &=& k\,\mathcal{W}(x, \, k;\,\tau)
,\\
\mathcal{J}_k(x, \, k;\,\tau) &=& -\sum_{\eta=0}^{\infty} \left(\frac{i}{2}\right)^{2\eta}\frac{1}{(2\eta+1)!} \, \left[\left(\frac{\partial~}{\partial x}\right)^{2\eta+1}\hspace{-.5cm}\mathcal{U}(x)\right]\,\left(\frac{\partial~}{\partial k}\right)^{2\eta}\hspace{-.3cm}\mathcal{W}(x, \, k;\,\tau),
\end{eqnarray}\normalsize
from which the continuity equation is given by
\begin{equation} 
\frac{\partial \mathcal{W}}{\partial \tau} + \frac{\partial \mathcal{J}_x}{\partial x}+\frac{\partial \mathcal{J}_k}{\partial k} = \frac{\partial \mathcal{W}}{\partial \tau} + \mbox{\boldmath $\nabla$}_{\xi}\cdot\mbox{\boldmath $\mathcal{J}$} =0.
\end{equation}

\section*{Appendix III - Low temperature limit for $\mathcal{W}^\alpha _ \Omega(x, \, k;\,\beta)$} \label{equaapp3}

By using the series expansion of the Bessel function the integral,
\begin{equation}
\begin{split}
\mathcal{W}^\alpha _ \Omega(x, \, k;\,\beta) & = \frac{2 \exp(\alpha \beta \hbar \omega) x^2}{\pi} \sum ^\infty _ {m=0} \frac{1}{\Gamma(m+1) \Gamma(m + \alpha + 1)} \left(\frac{x^2}{2\sinh(\beta \hbar \omega)}\right)^{2m + \alpha} \times \\ & \int ^{1} _{-1} d\sigma \exp[ 2 i k \sigma] (1 - \sigma^2)^{1/2 + 2m + \alpha} \exp[-\coth(\beta \hbar \omega) x^2 (1 + \sigma^2)],
\end{split}
\end{equation}
with $\sigma = y/x$, for half-integer values of $\alpha$, one has \[ (1 - \sigma^2)^{1/2 + 2m + \alpha} = \sum^{1/2 + \alpha + 2m} _{j=0} \frac{\Gamma(3/2 + \alpha + 2m)}{\Gamma(3/2 + \alpha + 2m - j) \Gamma(j+1)} (-1) ^j \sigma^{2j},\]
and thus
\begin{eqnarray}
\mathcal{W}^\alpha _ \Omega (x, \, k;\,\beta) &=& \left(\frac{\exp(\beta \hbar \omega) x^2}{2 \sinh(\beta \hbar \omega)} \right)^\alpha  \frac{x^2}{\pi ^{1/2}} \, \times \nonumber \\
&& \sum ^\infty _ {m=0} \frac{\Gamma(3/2 + \alpha + 2m)}{\Gamma(m+1) \Gamma(m + \alpha + 1)} \left( \frac{x^2}{2 \sinh(\beta \hbar \omega)}\right)^{2m}\mathcal{K}_m ^\alpha (x,k)
\end{eqnarray}
where
\begin{equation*}
\mathcal{K}_m ^\alpha (x,k) = \exp[-\zeta x^2] \sum^{1/2 + \alpha + 2m} _{j=0} \frac{d^j \zeta}{d \zeta^ {^j}} \left[ \zeta^{-1/2} \exp \left(-\frac{k^2}{\zeta}\right)2 \Re \left\{ \textrm{Erf}\bigg(\zeta^{1/2}(x + i \, \zeta^{-1}k) \bigg) \right\} \right] \bigg \rvert _{\zeta = \coth(\beta \hbar \omega)}.
\end{equation*}
One notices that for large values of $\sinh(\beta\hbar\omega)$, only the $m = 0$ term is effective.

\end{document}